\documentclass[runningheads]{llncs}
\usepackage{adjustbox}
\usepackage{multirow}
\usepackage{subcaption}
\usepackage{float}
\usepackage{hyperref}
\usepackage{url}
\usepackage{graphicx}
\usepackage{algorithm}
\usepackage[noend]{algpseudocode}
\usepackage{amssymb,amsmath, bm}
\usepackage{comment}
\usepackage{color}
\usepackage{booktabs}

\newcommand{\tabincell}[2]{\begin{tabular}{@{}c#1@{}}#2\end{tabular}} 
\hypersetup{
   colorlinks,
   menucolor=black,
   linkcolor=black,
   citecolor=red,
   urlcolor=blue
}

\begin{document}
%
\title{Defending Against Backdoor Attacks by Layer-wise Feature Analysis}
\author{Najeeb~Moharram~Jebreel\inst{1} \and
Josep Domingo-Ferrer\inst{1} \and Yiming~Li\inst{2}}
\authorrunning{Najeeb Jebreel et al.}
\institute{Universitat Rovira i Virgili\\
Av. Pa\"{\i}sos Catalans 26, E-43007 Tarragona, Catalonia\\
\email{\{najeeb.jebreel,josep.domingo\}@urv.cat} \and
Tsinghua University, Beijing, China\\
\email{li-ym18@mails.tsinghua.edu.cn}
}

\maketitle              
\begin{abstract}
Training deep neural networks (DNNs) usually requires massive training data and computational resources. Users who cannot afford this may prefer to outsource training to a third party or resort to publicly available pre-trained models. Unfortunately, doing so facilitates a new training-time attack ({\em i.e.}, backdoor attack) against DNNs. This attack aims to induce misclassification of input samples containing adversary-specified trigger patterns. In this paper, we first conduct a layer-wise feature analysis of poisoned and benign samples from the target class. We find out that the feature difference between benign and poisoned samples tends to be maximum at a critical layer, which is {\em not always} the one typically used in existing defenses, namely the layer before fully-connected layers. We also demonstrate how to locate this critical layer based on the behaviors of benign samples. We then propose a simple yet effective method to filter poisoned samples by analyzing the feature differences between suspicious and benign samples at the critical layer. We conduct extensive experiments on two benchmark datasets, which confirm the effectiveness of our defense.


\keywords{Backdoor Detection  \and Backdoor Defense \and Backdoor Learning \and AI Security \and Deep Learning.}
\end{abstract}

\section{Introduction}
\label{sec_introduction}
In recent years, deep neural networks (DNNs) 
have successfully been applied in many tasks, such as computer vision, natural language processing, and speech recognition. However, training DNNs requires massive training data and computational resources, and users who cannot afford it may opt to outsource training to a third-party ({\em e.g.}, a cloud service) or leverage pre-trained DNNs. 
Unfortunately, losing control over training facilitates  
 {\em backdoor attacks} \cite{chen2017targeted,gu2019badnets,li2022backdoor} against DNNs. 
In these attacks, the adversary poisons a few training samples to cause the DNN to misclassify samples 
containing pre-defined trigger patterns into an adversary-specified target class. Nevertheless, the attacked models behave normally on benign samples, which makes the attack stealthy. Since DNNs are used in many mission-critical tasks ({\em e.g.}, autonomous driving, or facial recognition), it is urgent to design effective defenses against these attacks. 


Among all backdoor defenses in the literature, backdoor detection is one of the most important defense paradigms, where defenders attempt to detect whether a suspicious object ({\em e.g.}, model or sample) is malicious. Currently, most existing backdoor detectors assume poisoned samples have different feature representations from benign samples, and they tend to focus on the layer before the fully connected layers \cite{chen2019detecting,tang2021demon,hayase2021spectre}. 
Two intriguing questions arise:
\textbf{(1)} \emph{Is this layer always the most critical place for backdoor detection?} 
\textbf{(2)} \emph{If not, how to find the critical layer for designing more effective backdoor detection?}

\begin{figure}[!t]
    \centering
      \includegraphics[width=0.8\linewidth]{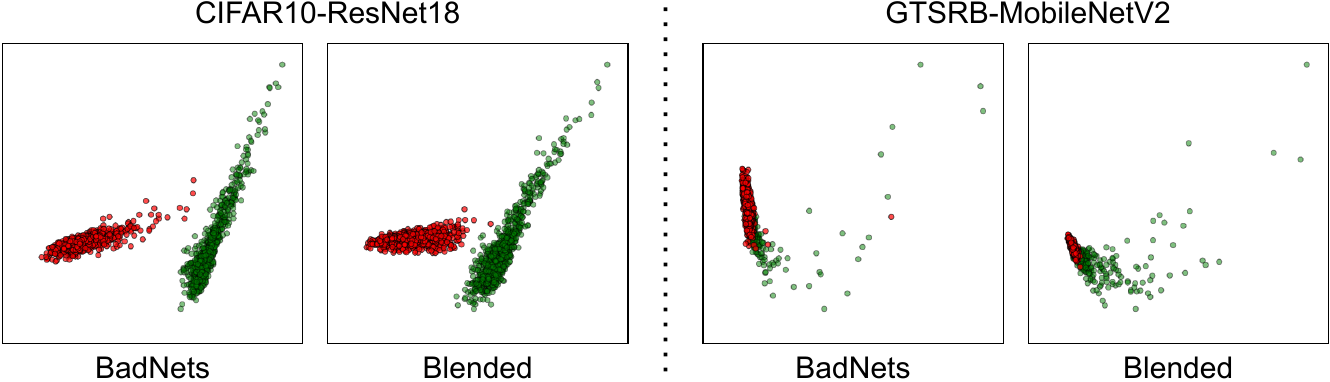}
\vspace{-0.4em}
\caption{PCA-based visualization of 
features of benign (green) and poisoned samples (red) generated by the layer before the fully connected layers of models attacked by BadNets \cite{gu2019badnets} and Blended \cite{chen2017targeted}. As shown in this figure, features of poisoned and benign samples are not well separable on the GTSRB benchmark.}
\vspace{-2em}
\label{fig:latenst_examples}
\end{figure}

In this paper, we give a negative answer to the first
question (see Figure \ref{fig:latenst_examples}). To answer the second one, we conduct a layer-wise feature analysis of poisoned and benign samples from the target class. We find out that the feature difference between benign and poisoned samples tends to reach the maximum at a critical layer, which can be easily located based on the behaviors of benign samples. Specifically, {\em the critical layer is the one or near the one that contributes most to assigning benign samples to their true class.} Based on this finding, we propose a simple yet effective method to filter poisoned samples by analyzing the feature differences (measured by cosine similarity) between incoming suspicious samples and a few benign samples at the critical layer. Our method can serve as a `firewall' for deployed DNNs to identify, block, and trace malicious inputs.
In short, our main contributions are four-fold. \textbf{(1)} We demonstrate that the features of poisoned and benign samples are not always clearly separable at the layer before fully connected layers, which is the one typically used in existing defenses. \textbf{(2)} We conduct a layer-wise feature analysis aimed at locating
the critical layer where the separation between poisoned and benign samples is neatest. \textbf{(3)} We propose a backdoor detection method to filter poisoned samples by analyzing the feature differences between suspicious and benign samples at the critical layer. \textbf{(4)} We conduct extensive experiments on two benchmark datasets to assess the effectiveness of our proposed defense.




\section{Related Work: Backdoor Attacks and Defenses}
\label{sec_related}

In this paper, we focus on backdoor attacks and defenses in image classification. Other deep learning tasks are out of our current scope.  

BadNets~\cite{gu2019badnets} was the first backdoor attack, which randomly selected a few benign samples and generated their poisoned versions by stamping a trigger patch onto their images and reassigning their label as the target label. 
Later \cite{chen2017targeted} noted that the poisoned image should be similar to its benign version for stealthiness; these authors proposed a blended attack by introducing trigger transparency. 
However, these attacks are with poisoned labels and therefore users can still detect them by examining the image-label relation. To circumvent this, \cite{turner2019label} proposed the clean-label attack paradigm, where the target label is consistent with the ground-truth label of poisoned samples. Specifically, in this paradigm, adversarial attacks were exploited to perturb the selected benign samples before conducting the standard trigger injection process. \cite{nguyen2020wanet} adopted image warping as the backdoor trigger, which modifies the whole image while preserving its main content. Besides, \cite{nguyen2020input} proposed the first sample-specific attack, where the trigger varies across samples. However, such triggers are visible and the adversaries need to control the whole training process. More recently, \cite{li2021invisible} introduced the first poison-only invisible sample-specific attack to address these problems. 

Existing backdoor defenses fall into three main categories: input filtering, input pre-processing, and model repairing.
{\bf Input filtering} intends to differentiate benign and poisoned samples based on their distinctive behaviors, like the separability of the feature representations of benign and poisoned samples. For example, \cite{hayase2021spectre} introduced a robust covariance estimation of feature representations to amplify the spectral signature of poisoned samples. \cite{zeng2021rethinking} proposed to filter inputs inspired by the understanding that poisoned images tend to have some high-frequency artifacts. \cite{gao2022design} proposed to blend various images on the suspicious one, since the trigger pattern can still mislead the prediction no matter what the background contents are. {\bf Input pre-processing} modifies each input sample before feeding it into the deployed DNN. Its rationale is to perturb potential trigger patterns and thereby prevent backdoor activation. \cite{liu2017neural} proposed the first defense in this category where they used an encoder-decoder to modify input samples. \cite{rosenfeld2020certified} employed randomized smoothing to generate a set of input neighbors and averaged their predictions. Further, \cite{li2021backdoor} demonstrated that if the location or appearance of the trigger is slightly different from that used for training, the attack effectiveness may degrade sharply. Based on this, they proposed to pre-process images with spatial transformations.
{\bf Model repairing} aims at erasing backdoors contained in the attacked DNNs. For example, \cite{liu2017neural,zhao2020bridging,li2021neural} showed that users can effectively remove backdoors by fine-tuning the attacked DNNs with a few benign samples. 
\cite{liu2018fine} revealed that model pruning can also remove backdoors effectively, because backdoors are mainly encoded in specific neurons. 
Very recently, \cite{zeng2022adversarial} proposed to repair compromised models with adversarial model unlearning.
In this paper, {\em we focus on input filtering,} which is very convenient to protect deployed DNNs.

\section{Layer-wise Feature Analysis}
\label{sec_analysis}

A deep neural network (DNN) $f(\bm{x})$ is composed by $L$ layers $f^l, l\in [1, L]$. Each $f^l$ has a weight matrix $\bm{w^l}$, a bias vector $\bm{b^l}$, and an activation function $\sigma^l$. The output of $f^l$ is $\bm{a^{l}} = f^l(\bm{a^{l-1}}) = \sigma^l(\bm{w^l} \cdot \bm{a^{l-1}} + \bm{b^l})$, where $f^1$ takes input $\bm{x}$ and $f^L$ outputs a vector $\bm{a^L}$ with $\mathcal{C}$ classes.
The vector $\bm{a^L}$ is softmaxed to get probabilities $\bm{p}$. A DNN has a feature extractor that maps $\bm{x}$ to latent features, which are input to fully connected layers for classification.

In this paper, we use DNNs as $\mathcal{C}$-class classifiers, where $y_i$ is the ground truth label of $\bm{x_i}$ and $\hat{y_i}$ is the index of the highest probability in $\bm{p_i}$. Also, activations of intermediate layers are analyzed for detecting poisoned samples.


We notice that the predictions of attacked DNNs for both benign samples from the target class and poisoned samples are all the target label. The attacked DNNs mainly exploit class-relevant features to predict these benign samples while they use trigger-related features for poisoned samples. We suggest that defenders could exploit this difference to design effective backdoor detection. To explore their main differences, we conduct a layer-wise analysis, as follows.

\begin{definition}[Layer-wise centroids of target class features] 
Let $f'$ be an attacked DNN with a target class $t$.
Let $X_t = \{\bm{x_{i}}\}_{i = 1}^{|X_t|}$ be benign samples with true class $t$, and let $\{\bm{a_{i}^1}, \ldots, \bm{a_{i}^L}\}_{i = 1}^{|X_t|}$ be their intermediate features generated  by $f'$.
The centroid of $t$'s benign features at layer $l$ is defined as $\bm{\hat{a}_t^l} =  \frac{1}{|X_t|}\sum_{i=1}^{|X_t|} \bm{a_{i}^l}$, and $\{\bm{\hat{a}_t^1}, \ldots, \bm{\hat{a}_t^L}\}$ is the set of layer-wise centroids of $t$'s benign features.
\end{definition}

\begin{definition}[Layer-wise cosine similarity] 
Let $\bm{a_j^l}$ be the features generated by layer $l$ for an input $\bm{x_j}$, and let $cs_{j}^l$ be the cosine similarity between $\bm{a_j^l}$ and the corresponding $t$'s centroid $\bm{\hat{a}_t^l}$.
The set $\{cs_{j}^1, \ldots, cs_{j}^L\}$ is said to be the layer-wise cosine similarities between $\bm{x_j}$ and $t$'s centroids.
\end{definition}

\noindent \textbf{Settings.} We conducted six representative attacks on four classical benchmarks: CIFAR10-ResNet18, CIFAR10-MobileNetV2, GTSRB-ResNet18, and GTSRB-MobileNetV2.
The six attacks were BadNets~\cite{gu2019badnets}, the backdoor attack with blended strategy (Blended) \cite{chen2017targeted}, the label-consistent attack (LC) of \cite{turner2019label}, WaNet~\cite{nguyen2020wanet}, ISSBA~\cite{li2021invisible}, and IAD~\cite{nguyen2020input}. More details on the datasets, DNNs, and attack settings are presented in Section~\ref{sec_experiments}.
Specifically, for each attacked DNN $f'$ with a target class $t$, we estimated $\{\bm{\hat{a}_t^1}, \ldots, \bm{\hat{a}_t^L}\}$ using $10\%$ of the benign test samples labeled as $t$. 
Then, for the benign and poisoned test samples classified by $f'$ into $t$, we calculated the layer-wise cosine similarities between their generated features and the corresponding estimated centroids.
Finally, we visualized the layer-wise means of the computed cosine similarities of the benign and poisoned samples to analyze their behaviors.

\begin{figure}[!t]
\vspace{-1em}
    \centering
    \begin{subfigure}{0.473\textwidth}
      \centering
      \includegraphics[width=1\linewidth]{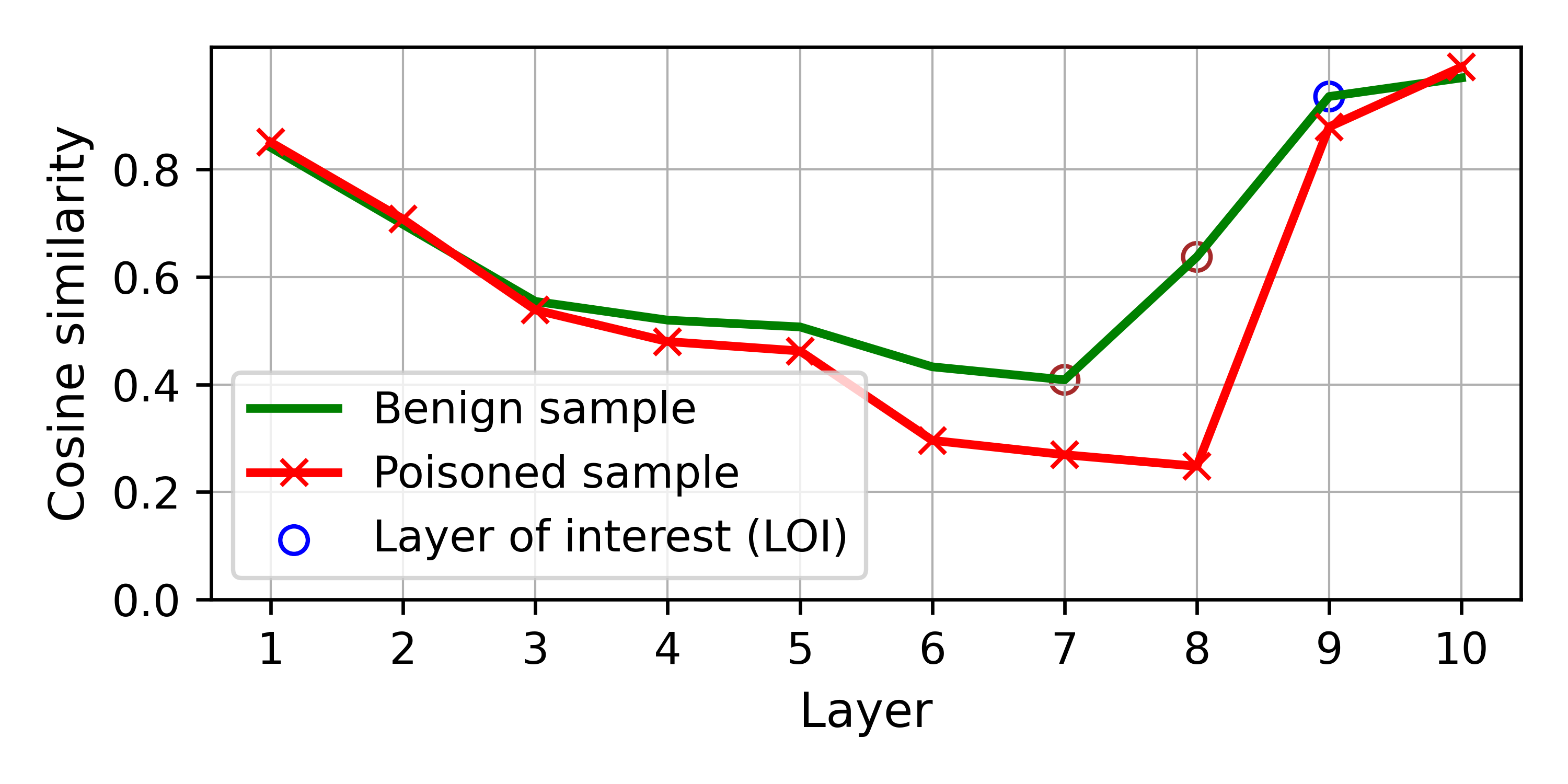}
      \vspace{-1.5em}
      \caption{BadNets}
    \end{subfigure}%
    \hspace{1.5em}
     \begin{subfigure}{0.473\textwidth}
      \centering
      \includegraphics[width=1\linewidth]{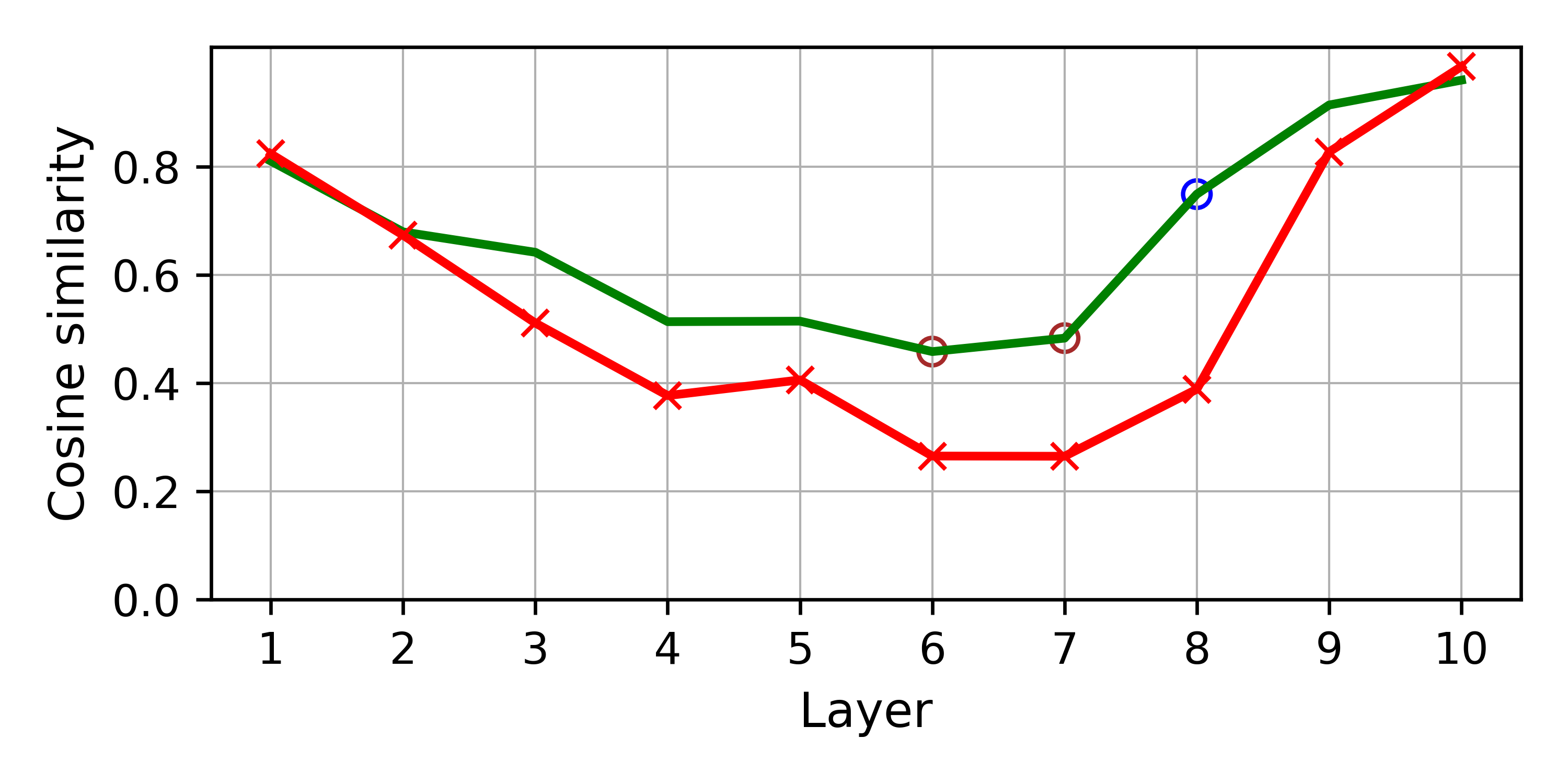}
      \vspace{-1.5em}
      \caption{ISSBA}
    \end{subfigure}
\vspace{-2em}
\caption{Layer-wise behaviors of benign samples from the target class and poisoned samples (generated by BadNets and ISSBA) on CIFAR-10 with ResNet-18}
\label{fig:cifar10_resnet18_badnet_issba_analysis}
\vspace{-0.1cm}
\end{figure}

\begin{figure}[!t]
    \centering
    \vspace{-1em}
    \begin{subfigure}{0.473\textwidth}
      \centering
      \includegraphics[width=1\linewidth]{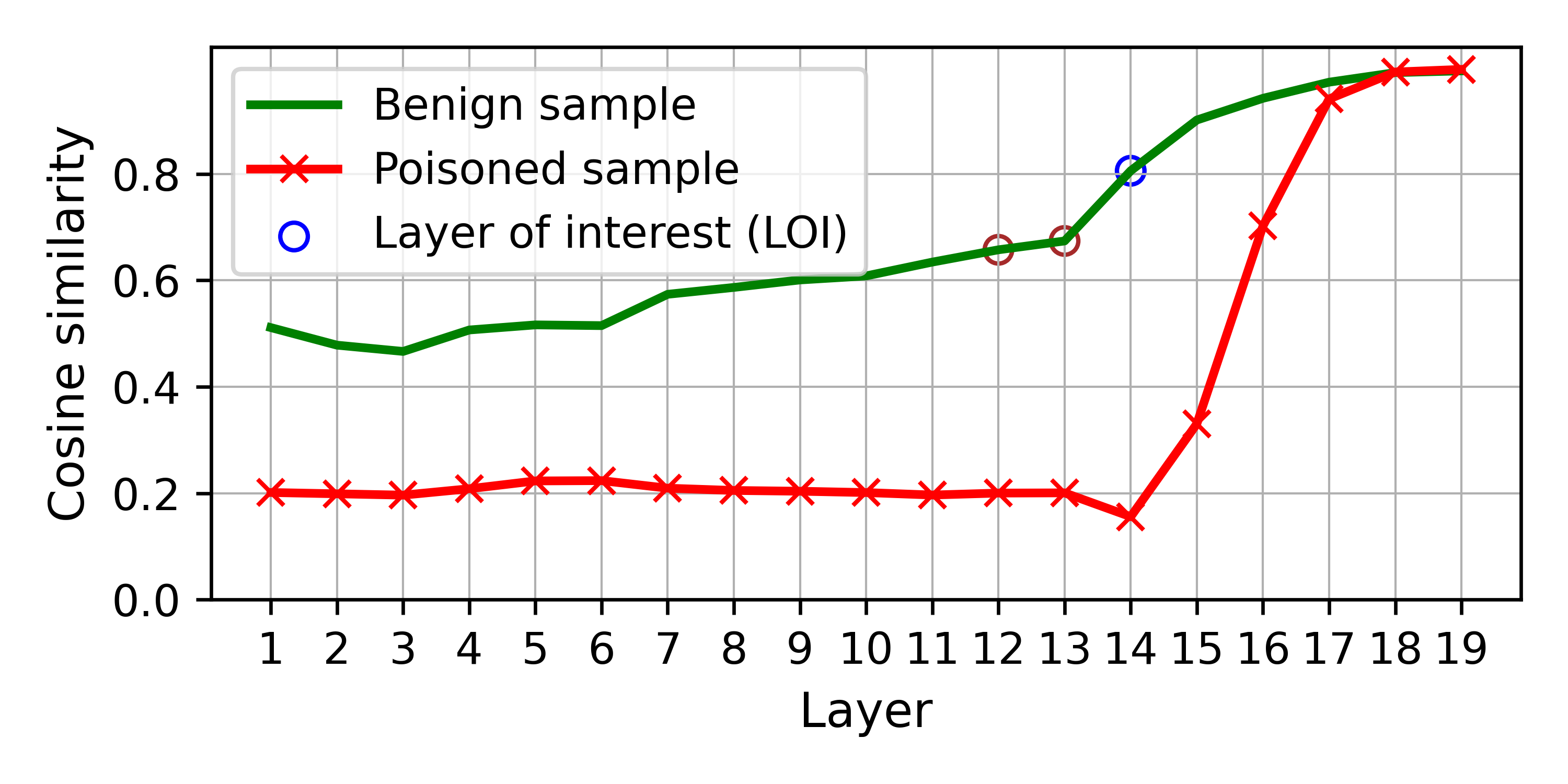}
      \vspace{-1.5em}
      \caption{BadNets}
    \end{subfigure}%
        \hspace{1.5em}
     \begin{subfigure}{0.473\textwidth}
      \centering
      \includegraphics[width=1\linewidth]{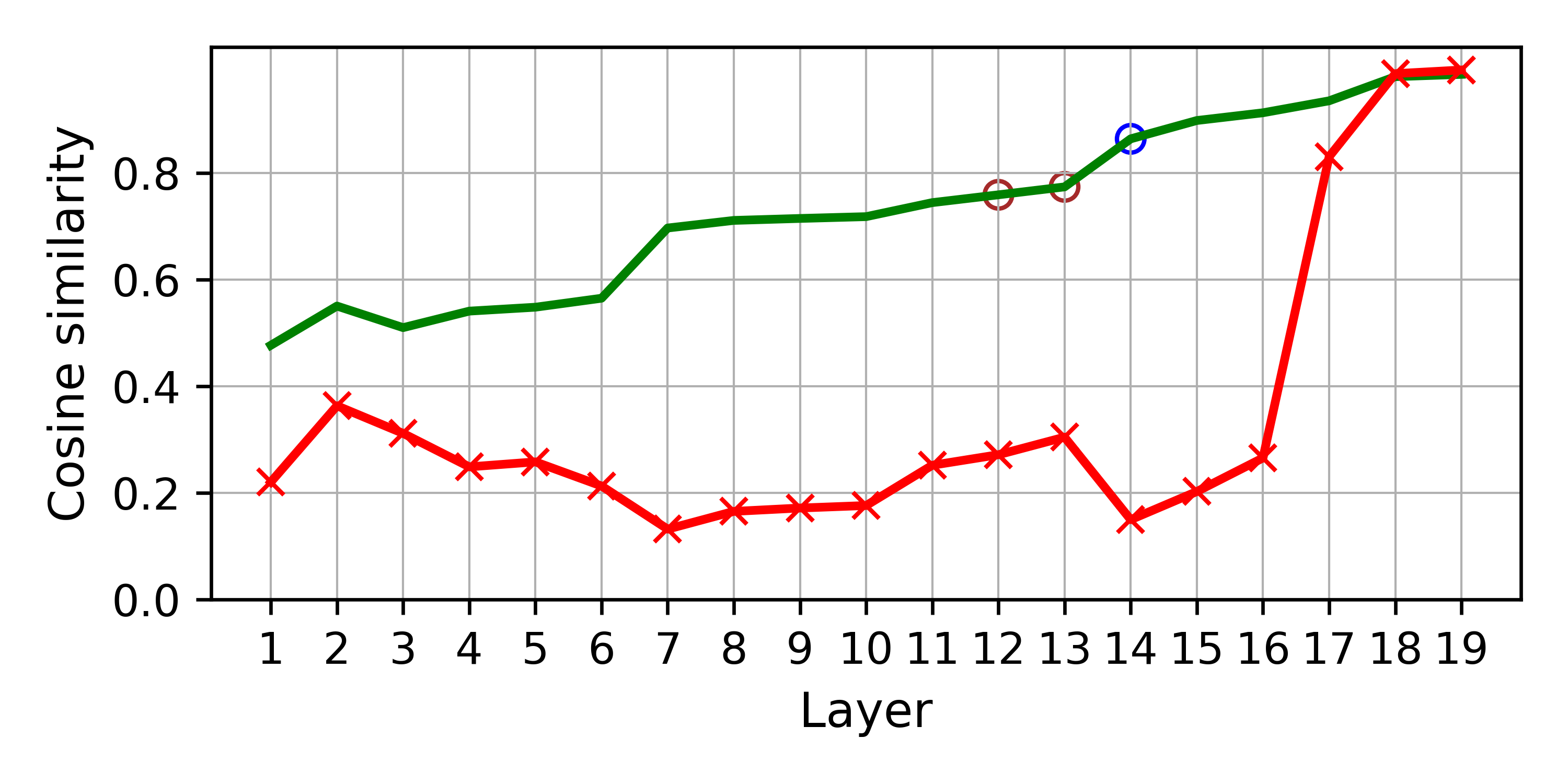}
      \vspace{-1.5em}
      \caption{ISSBA}
    \end{subfigure}
    \vspace{-0.2em}
\caption{Layer-wise behaviors of benign samples from the target class and poisoned samples (generated by BadNets and ISSBA) on GTSRB with MobileNetV2}
\label{fig:gtsrb_mobilenetv2_badnets_issba_analysis}
\vspace{-2em}
\end{figure}

\vspace{0.3em}
\noindent \textbf{Results.} Figure~\ref{fig:cifar10_resnet18_badnet_issba_analysis} shows the layer-wise means of cosine similarity for benign and poisoned samples with the CIFAR10-ResNet18 benchmark under the BadNets and ISSBA attacks.
As we go deeper into the attacked DNN layers, the gap between the direction of benign and poisoned features gets larger until we reach a specific layer where the backdoor trigger is activated, causing poisoned samples to get closer to the target class. 
Figure~\ref{fig:gtsrb_mobilenetv2_badnets_issba_analysis} shows the same phenomenon for the GTSRB-MobileNetV2 benchmark. 
Further, we can see that for BadNets
the latent features of benign and poisoned samples are similar in the last layer of the features extractor ($i.e.$, layer 17).

Regardless of the attack or benchmark, when we enter 
the second half of DNN layers (which usually are class-specific), {\em benign samples start to get closer to the target class before the poisoned ones, that are still farther from the target class} because the backdoor trigger is not yet activated. This makes the difference in similarity maximum in one of those latter layers, which we call the {\em critical layer}. In particular, \emph{this layer is not always the one typically used in existing defenses} ({\em i.e.}, the layer before fully-connected layers). Besides, we show that it is very likely to be either the layer that contributes most to assigning the benign samples to their true target class (which we name the {\em layer of interest or LOI}, circled in blue) or one of the two layers before the LOI (circled in brown). 

Results under other attacks for these benchmarks are presented in Appendix~\ref{sec_more_analysis}.
In those materials, we also provide confirmation that the above
distinctive behaviors hold regardless of the datasets or models being used.
From the analysis above, we can conclude that focusing on those circled layers
can help develop a simple and robust defense against backdoor attacks.

\section{The Proposed Defense}
\label{sec_meth}
\noindent \textbf{Threat Model.} Consider a user that obtains a suspicious trained $f_s$ that might contain hidden backdoors. 
We assume that the user has limited computational resources or benign samples, and therefore cannot repair $f_s$. The user wants to defend by detecting at inference time whether a suspicious incoming input $\bm{x_s}$ is poisoned, given $f_s$.  
Similar to existing defenses, we assume that a small set of benign samples $X_{val}$ is available to the user/defender. We denote the available samples that belong to a potential class $t$ as $X_{t_{val}}$. Let $m = |X_{t_{val}}|$ denote the number of available samples labeled as $t$.

\vspace{0.3em}
\noindent {\bf Method Design.} 
Based on the lessons learned in Section~\ref{sec_analysis}, our method 
to detect poisoned samples at inference time consists of four steps.
\textbf{1)} Estimate the layer-wise features' centroids of class $t$ for each
of layers $\lfloor L/2\rfloor$ to $L$ using the class's available benign samples.
\textbf{2)} Compute the cosine similarities between the extracted features and the estimated centroids, and then compute the layer-wise means of the computed cosine similarities.
\textbf{3)} Identify the layer of interest (LOI)
as per Algorithm~\ref{alg1}, sum up the cosine similarities in LOI and the two layers before LOI (sample-wise), and compute the mean and standard deviation of the summed cosine similarities.
\textbf{4)} For any suspicious incoming input $\bm{x_s}$ classified as $t$ by $f_s$, \textbf{4.1)} compute its cosine similarities to the estimated centroids in the above-mentioned three layers, and \textbf{4.2)} consider it as a potentially poisoned input if its summed similarities fall below the obtained mean by a specific number $\tau$ of standard deviations (called threshold in what follows).
A detailed pseudocode can be found in Appendix~\ref{sec_meth_detailed}.

\begin{algorithm}[t!]
\caption{Identify layer of interest (LOI).}\label{alg1}
\hspace*{\algorithmicindent} \textbf{Input}: Cosine similarities $\{\hat{cs}^{\lfloor L/2 \rfloor}_t,\ldots, \hat{cs}^{L}_t\}$ for potential target class $t$
\begin{algorithmic}[1]
\State $max_{diff}\gets\hat{cs}^{\lfloor L/2 \rfloor + 1}_t - \hat{cs}^{\lfloor L/2\rfloor}_t$; $LOI_t \gets\lfloor L/2\rfloor + 1$;
        \For{\texttt{$l\in\{\lfloor L/2 \rfloor  + 2, \ldots, L\}$}}
            \State $l_{diff}\gets\hat{cs}^{l}_t - \hat{cs}^{l-1}_t$;
            \If{$l_{diff} > max_{diff}$}
                \State $max_{diff}\gets l_{diff}$; $LOI_t\gets l$;
            \EndIf
         \EndFor
\State {\bf return} $LOI_t$.
\end{algorithmic}
\end{algorithm}

\begin{figure}[!t]
    \centering
    \vspace{-0.6em}
\includegraphics[width=0.88\linewidth]{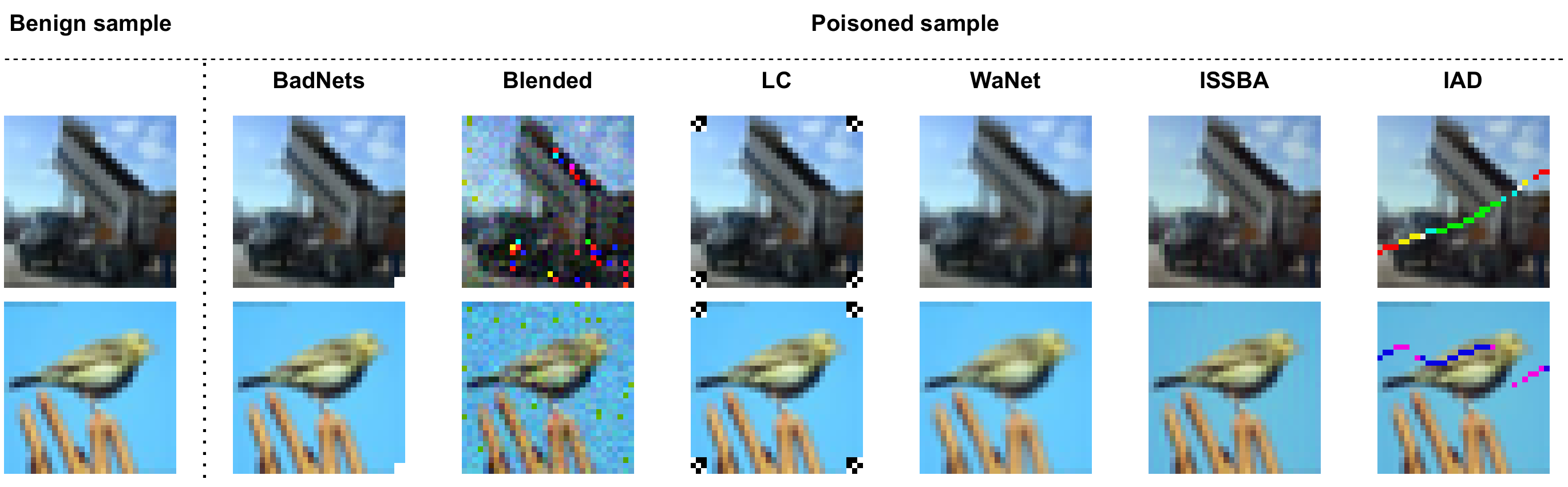}
\vspace{-0.6em}
\caption{The example of benign samples and their poisoned versions generated by six representative backdoor attacks.}
\label{fig:attacks_example}
\vspace{-1.5em}
\end{figure}

\section{Experiments}
\label{sec_experiments}
\vspace{-0.5em}
\subsection{Main Settings}
\vspace{-0.5em}
\noindent \textbf{Datasets and DNNs.} In this paper, we use two classic benchmark datasets, namely CIFAR10~\cite{krizhevsky2009learning} and GTSRB~\cite{stallkamp2011german}. We use the ResNet18~\cite{he2016deep} on CIFAR10 and the MobileNetV2~\cite{sandler2018mobilenetv2} on GTSRB. More details are presented in Appendix~\ref{sec_datasets_dnns}. 
The source code, pre-trained models, and poisoned test sets of our defense are available at \url{https://github.com/NajeebJebreel/DBALFA}.

\vspace{0.3em}
\noindent \textbf{Attack Baselines.}
We evaluated each defense under the six attacks mentioned in
Section~\ref{sec_analysis}: BadNets, Blended, LC, WaNet, ISSBA, and IAD. They are representative of visible attacks, patch-based invisible attacks, clean-label attacks, non-patch-based invisible attacks, invisible sample-specific attacks, and visible sample-specific attacks, respectively.

\vspace{0.3em}
\noindent \textbf{Defense Baselines.}
We compared our defense with six representative defenses, namely randomized smoothing (RS) \cite{rosenfeld2020certified}, ShrinkPad (ShPd) \cite{li2021backdoor}, activation clustering (AC) \cite{chen2019detecting}, STRIP \cite{gao2022design}, SCAn \cite{tang2021demon}, and fine-pruning (FP) \cite{liu2018fine}. RS and ShPd are two defenses with input pre-processing; AC, STRIP, and SCAn are three advanced input-filtering-based defenses; FP is based on model repairing.

\vspace{0.3em}
\noindent \textbf{Attack Setup.}
For both CIFAR10 and GTSRB, we took the following settings.
We used a $2\times2$ square as the trigger pattern for BadNets (as suggested in~\cite{gu2019badnets}). We adopted the random noise pattern, with a $10\%$ blend ratio, for Blended (as suggested in~\cite{chen2017targeted}). 
The trigger pattern adopted for the LC attack was the same used in BadNets. 
For WaNet, ISSBA, and IAD, we took their default settings. 
Besides, we set the poisoning rate to $5\%$ for BadNets, Blended, LC, and ISSBA. 
For WaNet and IAD, we set the poisoning rate to $10\%$. We implement baseline attacks based on the codes in \texttt{BackdoorBox} \cite{li2023backdoorbox}. More details on settings are given in Appendix~\ref{sec_attack_setting}.
Figure~\ref{fig:attacks_example} shows an example of poisoned samples generated by different attacks.

\vspace{0.3em}
\noindent \textbf{Defense Setup. }
For RS, ShPd and STRIP, we took the settings suggested in~\cite{rosenfeld2020certified,li2021backdoor,gao2022design}.
For FP, we pruned $95\%$ of the dormant neurons in the last convolution layer and fine-tuned 
the pruned model using $5\%$ of the training set.
We adjusted RS, ShPd, and FP to be used as detectors for poisoned samples by comparing the prediction change before and after applying them to an incoming input.
For AC, STRIP, SCAn, and our defense, we randomly selected $10\%$ from each benign test set 
as the available benign samples.
For SCAn, we identified classes with scores larger than $e$ as potential target classes, as suggested in~\cite{tang2021demon}.
For our defense, we used a threshold $\tau = 2.5$, which gives a reasonable trade-off between TPR and FPR for both benchmarks.

\vspace{0.3em}
\noindent \textbf{Evaluation Metrics. }
We used the main accuracy (MA) and the attack success rate (ASR) to measure attack performance.
Specifically, MA is the number of correctly classified benign samples divided by the total number of benign samples, and ASR is the number of poisoned samples classified as the target class divided by the total number of poisoned samples.
We adopted TPR and FPR to evaluate the performance of all defenses, where
TPR is computed as the number of detected poisoned inputs divided by the total number of poisoned inputs, whereas FPR is the number of benign inputs falsely detected as poisoned divided by the total number of benign inputs.

\begin{table}[!t]
\centering
\caption{Main results (\%) on the CIFAR-10 dataset. Boldfaced values are the best results among all defenses. Underlined values are the second-best results.}
\label{tab:cifar10_resnet18_defenses}
\resizebox{\textwidth}{!}{%
\begin{tabular}{c|cccccccccccc|cc}
\toprule
\multicolumn{1}{c|}{Attack$\rightarrow$}         & \multicolumn{2}{c}{BadNets}        & \multicolumn{2}{c}{Blended}     & \multicolumn{2}{c}{LC}          & \multicolumn{2}{c}{WaNet}      & \multicolumn{2}{c}{ISSBA}      & \multicolumn{2}{c|}{IAD}       & \multicolumn{2}{c}{Avg}        \\ \hline
\multicolumn{1}{c|}{\tabincell{c}{Metric$\rightarrow$\\Defense$\downarrow$}} & TPR          & FPR         & TPR           & FPR        & TPR          & FPR         & TPR         & FPR        & TPR         & FPR         & TPR          & FPR         & TPR          & FPR         \\ \hline
RS                                  & 9.84           & 8.00          & 7.35           & 5.76          & 9.21            & 7.52          & 98.48          & 10.00         & 8.83           & 8.72          & 13.28          & 6.36          & 24.50          & 7.73          \\
ShPd                                & 94.28          & 13.31         & 49.72           & 12.89         & 69.87           & 13.18         & 36.25          & 17.69         & 95.22          & 5.50          & 42.74          & 7.56          & 64.68          & 11.69         \\
FP                                  & 96.10          & 17.13         & \underline{96.23}           & 16.16         & \underline{94.76}           & 17.31         & 96.01          & 18.64         & 98.98          & 19.53         & \underline{97.08}          & 22.52         & 96.53          & 18.55         \\
AC                                  & \textbf{99.52} & 31.14         & \textbf{100.00}          & 30.69         & \textbf{100.00} & 31.16         & \textbf{99.18} & 32.44         & \textbf{99.94} & 34.22         & 82.99          & 31.32         & \underline{96.94}          & 31.83         \\
STRIP                               & 68.70          & 11.70         & 65.20           & 11.70         & 66.00           & 12.80         & 7.90           & 12.30         & 56.20          & 11.40         & 2.10           & 14.00         & 44.35          & 12.32         \\
SCAn                                & 96.60          & \textbf{0.77} & \textbf{100.00} & \textbf{0.00} & 0.02            & \underline{5.05}          & \underline{98.55}          & \textbf{1.06} & \underline{99.89}          & \underline{2.61}          & 84.19          & \textbf{0.13} & 79.88          & \underline{1.60}          \\ \hline
Ours                                & \underline{99.38}          & \underline{1.35}          & \textbf{100.00} & \underline{1.59}          & \textbf{100.00} & \textbf{1.20} & 91.04          & \underline{1.48}          & 98.97          & \textbf{1.17} & \textbf{99.12} & \underline{1.26}          & \textbf{98.09} & \textbf{1.34} \\ \bottomrule
\end{tabular}
}
\vspace{-0.5cm}
\end{table}

\begin{table}[!t]
\centering
\caption{Main results (\%) on the GTSRB dataset. 
Boldfaced values are the best results among all defenses. Underlined values are the second-best results.}
\label{tab:gtsrb_mobilenetv2_defenses}
\resizebox{\textwidth}{!}{%
\begin{tabular}{c|cccccccccccc|cc}
\toprule
\multicolumn{1}{c|}{Attack$\rightarrow$}         & \multicolumn{2}{c}{BadNets}        & \multicolumn{2}{c}{Blended}     & \multicolumn{2}{c}{LC}          & \multicolumn{2}{c}{WaNet}      & \multicolumn{2}{c}{ISSBA}      & \multicolumn{2}{c|}{IAD}       & \multicolumn{2}{c}{Avg}        \\ \hline
\multicolumn{1}{c|}{\tabincell{c}{Metric$\rightarrow$\\Defense$\downarrow$}} & TPR          & FPR         & TPR           & FPR        & TPR          & FPR         & TPR         & FPR        & TPR         & FPR         & TPR          & FPR         & TPR          & FPR         \\ \hline
RS                                  & 13.20          & 22.10         & 10.12           & 20.40         & 9.23            & 19.15         & 10.10           & 17.20         & 8.61            & 16.98         & 17.70           & 17.60         & 11.49           & 18.91                     \\
ShPd                                & \underline{94.97}          & 12.16         & 11.58           & 10.68         & \underline{96.16}           & 10.60         & 66.11           & 14.81         & 95.92           & 8.26          & 31.07           & 16.10         & 65.97           & 12.10                     \\
FP                                  & 89.05          & 18.80         & 30.56           & \textbf{3.70} & 94.71           & 50.02         & \underline{67.12}           & \underline{3.24}          & 94.22           & 7.05          & \underline{94.37}           & \underline{5.75}          & \underline{78.34}           & 14.76                     \\
AC                                  & 0.30           & 8.84          & 0.00            & \underline{5.67}          & 4.83            & \textbf{5.42} & 0.42            & 25.87         & \underline{99.06}           & 17.48         & 43.85           & 10.73         & 24.74           & 12.34                     \\
STRIP                               & 32.00          & 9.00          & \underline{80.40}           & 10.80         & 7.40            & 11.00         & 34.20           & 11.40         & 13.00           & 13.60         & 6.60            & 10.60         & 28.93           & 11.07                     \\
SCAn                                & 46.05          & \textbf{2.57} & 46.02           & 4.03          & 30.45           & 11.39         & 54.07           & \textbf{1.88} & 96.85           & \textbf{0.17} & 0.09            & 19.41         & 45.59           & \underline{6.58}                      \\ \hline
Ours                                & \textbf{99.99} & \underline{6.23}          & \textbf{100.00} & 6.72          & \textbf{100.00} & \underline{5.95}          & \textbf{100.00} & 6.49          & \textbf{100.00} & \underline{5.43}          & \textbf{100.00} & \textbf{4.67} & \textbf{100.00} & \textbf{5.92}             \\ \bottomrule
\end{tabular}}
\vspace{-0.5cm}
\end{table}

\vspace{-0.5em}
\subsection{Main Results}
\vspace{-0.2em}
For each attack, we ran each defense five times for a fair comparison. Due to space limitations, we present the average TPR and FPR in this section. 
Please refer to Appendix~\ref{sec_ablation_studies} for more detailed results.

As shown in Tables \ref{tab:cifar10_resnet18_defenses} and \ref{tab:gtsrb_mobilenetv2_defenses}, existing defenses failed to detect attacks with low TPR or high FPR in many cases, especially on the GTSRB dataset. For example, AC failed in most cases on GTSRB, although it had promising performance on CIFAR-10. In contrast, our method had good performance in detecting all attacks on both datasets. There were 
only a few cases (4 over 28) where our approach was neither optimal nor close to optimal. In these cases, our detection was still on par with state-of-the-art methods, and another indicator ({\em i.e.}, TPR or FPR) was significantly better than them. For example, when defending against the blended attack on the GTSRB dataset, the TPR of our method was 69.44\% larger than that of FP, which had the smallest FPR in this case. These results confirm the effectiveness of our detection.

\vspace{-0.3em}
\subsection{Discussions}
\vspace{-0.5em}
\noindent \textbf{Performance of Attacks.}
\label{sec_attack_performance}
Table~\ref{tab:attacks_results} shows the performance of the selected attacks on the CIFAR10-ResNet18 and the GTSRB-MobileNetV2 benchmarks.
It can be seen that sample-specific attacks ({\em e.g.}, ISSBA and IAD) performed better than other attacks in terms of MA and ASR.
\begin{table}[t!]
\centering
\caption{MA\% and ASR\% under the selected backdoor attacks on the CIFAR10-ResNet18 and the GTSRB-MobileNetV2 benchmarks. Best scores are in bold.}
\label{tab:attacks_results}
\resizebox{\textwidth}{!}{%
\begin{tabular}{c|c|cccccc}
\hline
\multicolumn{1}{c|}{Benchmark$\downarrow$}     & Metric$\downarrow$,Attack$\rightarrow$ & BadNets & Blended        & LC             & WaNet & ISSBA          & IAD            \\ \hline
\multirow{2}{*}{CIFAR10-ResNet18}  & MA\%          & 91.45   & 92.19          & 91.98          & 91.13 & \textbf{94.74} & 94.42          \\
                                   & ASR\%         & 97.20   & \textbf{100.0} & 99.96          & 99.04 & \textbf{100.0} & 99.66          \\ \hline
\multirow{2}{*}{GTSRB-MobileNetV2} & MA\%          & 97.00   & 97.27          & 97.45          & 96.09 & 98.43          & \textbf{98.81} \\
                                   & ASR\%         & 95.49   & \textbf{100.0} & \textbf{100.0} & 91.82 & \textbf{100.0} & 99.63          \\ \hline
\end{tabular}
}
\end{table}

\begin{figure}[t!]
    \centering
    \begin{subfigure}{0.48\textwidth}
      \centering
      \includegraphics[width=1\linewidth]{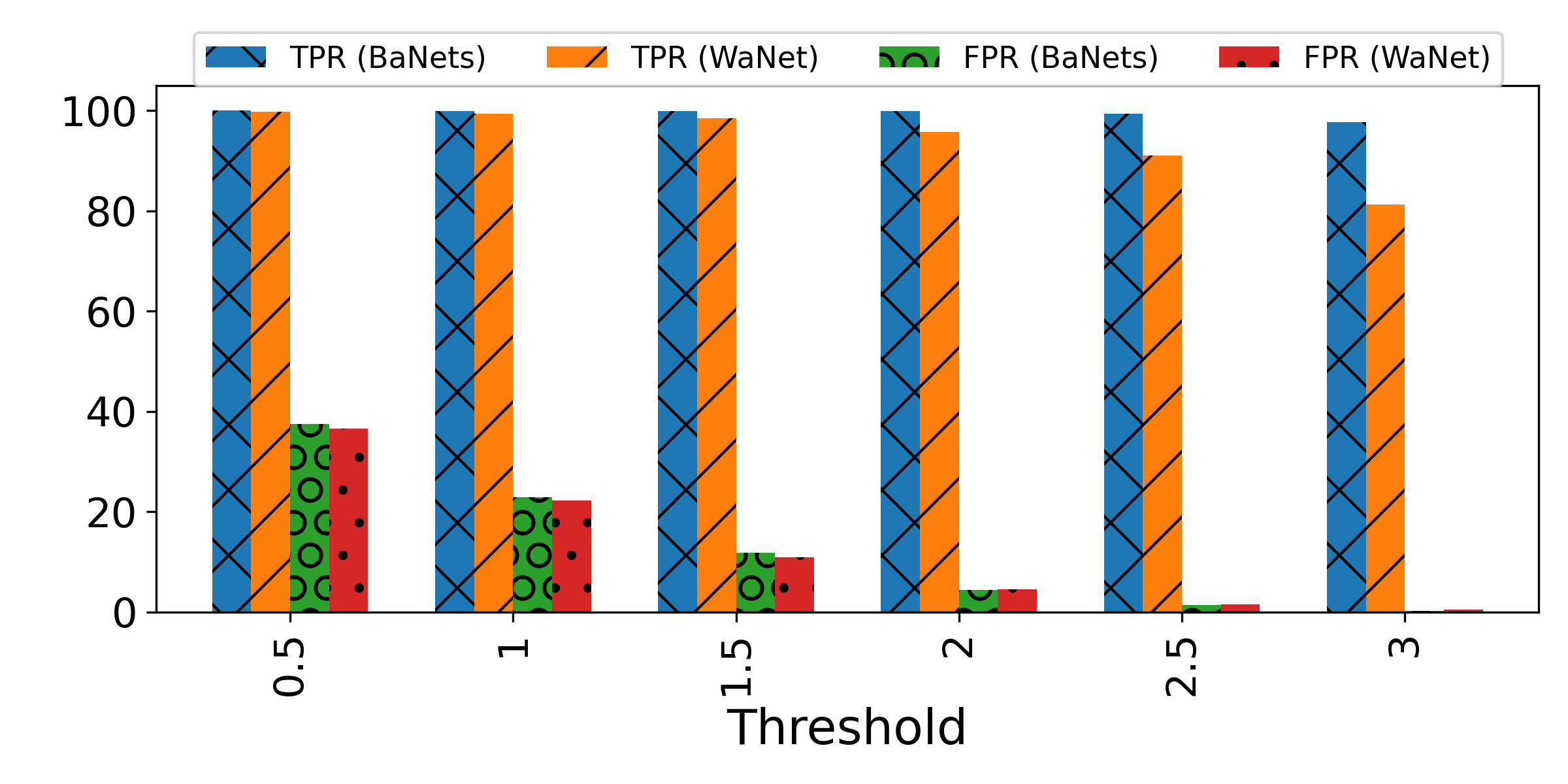}
      \vspace{-1em}
      \caption{CIFAR10-ResNet18}
    \end{subfigure}\hspace{1em}
    \begin{subfigure}{0.48\textwidth}
      \centering
      \includegraphics[width=1\linewidth]{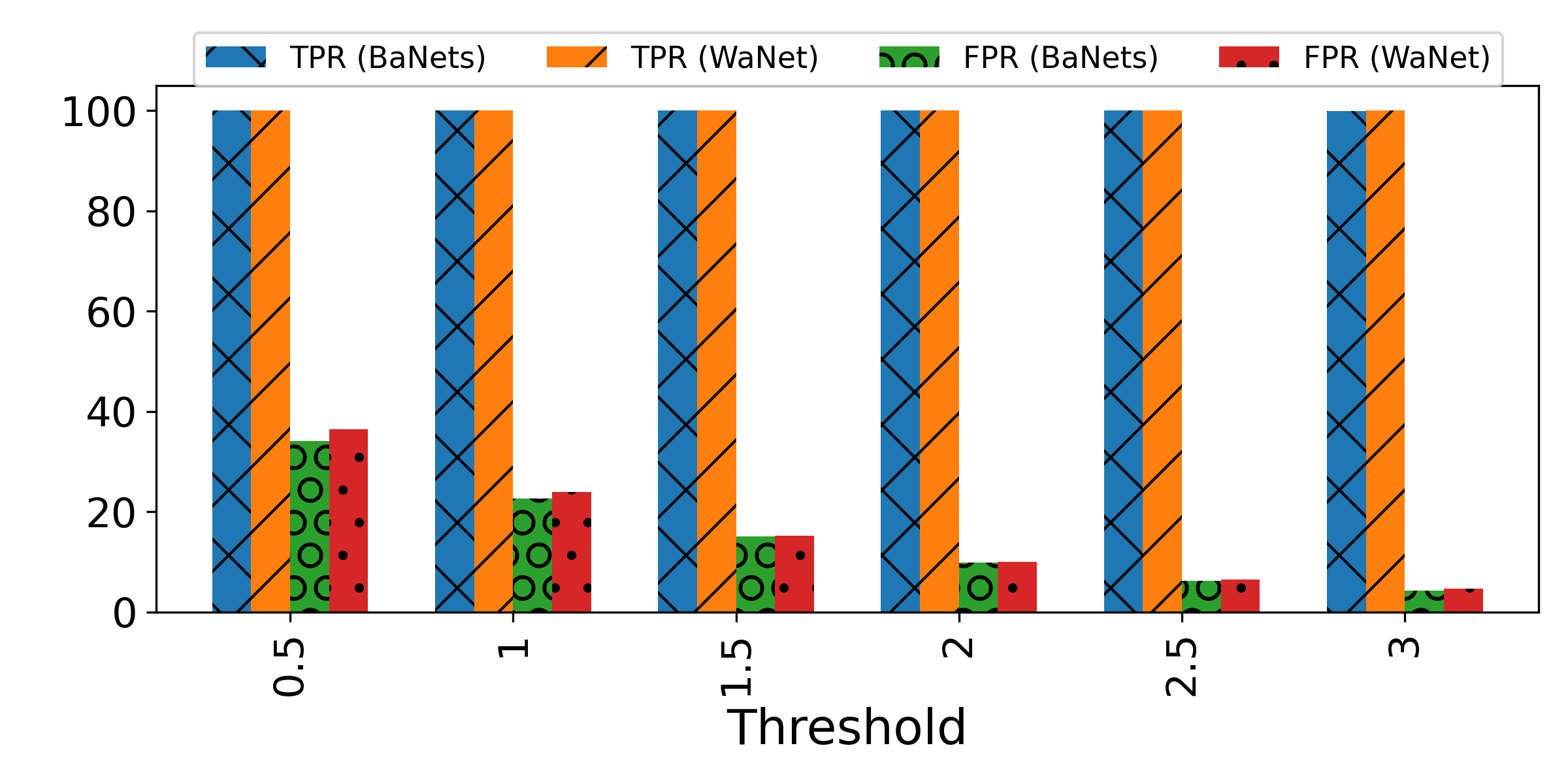}
      \vspace{-1em}
      \caption{GTSRB-MobileNetV2}
    \end{subfigure}%
    \vspace{-0.6em}
\caption{Impact of detection thresholds on TPR (\%) and FPR (\%)}
\label{fig:threshold_impact}
\vspace{-0.5cm}
\end{figure}

\vspace{0.3em}
\noindent \textbf{Effects of the Detection Threshold.} 
Figure \ref{fig:threshold_impact} shows the TPRs and FPRs of our defense with threshold  $\tau \in \{0.5, 1, 1.5, 2, 2.5, 3\}$ for BadNets and WaNet. It can be seen that a threshold 2.5 is reasonable, as it offers a high TPR while keeping a low FPR. Note that the larger the threshold, the smaller the TPR and FPR. Users should choose the threshold based on their specific needs.

\begin{table}[t!]
\centering
\caption{Impact of poisoning rates}
\label{tab:poisoning_rate_impact}
\begin{tabular}{c|llll}
\toprule
Poisoning Rate$\downarrow$, Metric$\rightarrow$ & MA (\%)           & ASR (\%)          & TPR (\%)          & FPR (\%)         \\ \hline
1\%                    & 91.52          & 94.15          & 99.64          & 1.25 \\
3\%                    & 92.28 & 96.31          & 99.32          & 1.32          \\
5\%                    & 91.45          & 97.20          & 99.36          & 1.35          \\
10\%                   & 91.45          & 97.56 & 99.83 & 1.62          \\ \bottomrule
\end{tabular}
\vspace{-0.5cm}
\end{table}

\begin{table}[t!]
\centering
\caption{Effectiveness of defenses with different features. Latent features denote those generated by the feature extractor that is typically used in existing defenses. Critical features are extracted by our method from the identified layers.}
\label{tab:critical_layer_impact}
\scalebox{0.95}{
\begin{tabular}{c|cc|cc}
\toprule
Metric$\rightarrow$                        & \multicolumn{2}{c}{TPR (\%)}                            & \multicolumn{2}{c}{FPR (\%)}                           \\ \hline
Defense$\downarrow$, Features$\rightarrow$ & Latent Features                       & Critical Features                        & Latent Features                       & Critical Features                       \\ \hline
AC                                         & 0.3                      & \textbf{96.32}                     & 8.84                     & \textbf{7.67}                     \\ \hline
SCAn                                       & 46.05                    & \textbf{86.19}                     & 2.57                     & \textbf{1.96}                     \\ \hline
Ours                      & 1.31 & \textbf{99.99} & \textbf{4.93} & 6.23 \\ \bottomrule
\end{tabular}
}
\vspace{-0.5cm}
\end{table}

\vspace{0.3em}
\noindent \textbf{Effects of the Poisoning Rate.} We launched BadNets on CIFAR10-ResNet18 using different poisoning rates $\in \{1\%, 3\%, 5\%, 10\%\}$ to study the impact of poisoning rates on our defense. Table~\ref{tab:poisoning_rate_impact} shows the attack success rate (ASR) increases with the poisoning 
rate. However, the poisoning rate has minor effects on our TPR and FPR. These results confirm
again the effectiveness of our method.

\vspace{0.3em}
\noindent \textbf{Effectiveness of Our Layer Selection.}
We compared the performance of AC, SCAn, and our method at 
detecting BadNets on the GTSRB-MobileNetV2 benchmark using latent features and critical features. 
We generated latent features based on the feature extractor ({\em i.e.}, the layer before fully-connected layers) that is typically adopted in existing defenses. The critical features were extracted by the layer of interest (LOI) used in our method. Table~\ref{tab:critical_layer_impact} shows that using 
our features led to significantly better performance in almost all cases. 
In other words, existing detection methods can also benefit from our LOI selection. 
Also, we compared the performance of our method on CIFAR10-ResNet18 under WaNet and IAD when using the features of every individual layer, and when using LOI and the two layers before LOI. 
Table~\ref{tab:individual_layer_performance} shows that as we approach the critical layer, which was just before LOI with WaNet and at LOI with IAD, the detection performance gets better. Since our method included the critical layer, it also was effective. 
These results confirm the effectiveness of our layer selection and partly explain our method's good
performance.

\begin{table}[t!]
\centering
\caption{Performance of features from individual layers compared to identified layers by our defense. The LOI of WaNet and IAD are 9 and 8, respectively.}
\label{tab:individual_layer_performance}
\begin{tabular}{cc|ccccccccccc} 
\toprule
\multicolumn{2}{c|}{Layer}        & 1     & 2     & 3    & 4     & 5     & 6     & 7     & 8     & 9     & 10   & Ours      \\ 
\hline
\multirow{2}{*}{WaNet} & TPR (\%) & 0.00  & 0.10  & 0.05 & 0.00  & 0.01  & 0.00  & 68.82 & 98.08 & 59.82 & 0.00 & 91.04     \\
                       & FPR (\%) & 0.09  & 0.82  & 0.24 & 0.20  & 0.21  & 0.04  & 2.06  & 1.52  & 2.06  & 0.65 & 1.48      \\ 
\hline
\multirow{2}{*}{IAD}   & TPR (\%) & 19.32 & 34.03 & 6.44 & 30.49 & 61.09 & 78.65 & 88.81 & 99.65 & 99.10 & 2.36 & ~ ~99.12  \\
                       & FPR (\%) & 1.65  & 1.38  & 1.44 & 1.60  & 2.27  & 1.70  & 1.29  & 1.13  & 1.09  & 1.24 & 1.26      \\
\bottomrule
\end{tabular}
\vspace{-0.5cm}
\end{table}

\vspace{0.3em}
\noindent \textbf{Effectiveness of Cosine Similarity.}
We compared the cosine similarity with the Euclidean distance as 
a metric to differentiate between benign and poisoned samples.
In Appendix~\ref{sec_cs_effectivness}, we show the cosine similarity gives 
a better differentiation than the Euclidean distance.
This is mostly because the direction of features is more important for detection 
than their magnitude.

\vspace{0.3em}
\noindent \textbf{Resistance to Adaptive Attacks.}
The adversary may adapt his attack to bypass our defense by optimizing the model's 
original loss $\mathcal{L}_{org}$ and minimizing the layer-wise angular deviation 
between the features of the poisoned samples and the features' centroids of the 
target class's benign samples. We studied the impact of this strategy by 
introducing the \emph{cosine distance} between the features of poisoned samples 
and the target class centroids as a secondary loss function $\mathcal{L}_{cd}$ in the training objective function. Also, we introduced a penalty parameter $\beta$, which yielded a 
modified objective function  $(1 - \beta)\mathcal{L}_{org} + \beta \mathcal{L}_{cd}$.
The role of $\beta$ is to control the trade-off between the angular deviation and the main accuracy loss.
We then launched BadNets on CIFAR10-ResNet18 under the modified objective function.
Table~\ref{tab:penalty_impact} (top subtable) 
shows MA and ASR with different penalty factors.
We can see that values of $\beta <0.9$  slightly increased the main accuracy because the 
second loss acted as a regularizer to the model's parameters, which reduced over-fitting. Also, 
 ASR stayed similar to the non-adaptive ASR (when $\beta = 0$).
However, the main accuracy degraded with greater $\beta$ values, because the original loss function was dominated by the angular deviation loss.
\begin{table}[!t]
\centering
\caption{Adaptive attack. Top, impact of penalty factor $\beta$ on MA and ASR.
Bottom, impact of penalty factor $\beta$ on TPR and FPR.}
\label{tab:penalty_impact}
\begin{tabular}{c|cccccclcc} 
\toprule
$\beta$  & 0     & 0.5   & 0.6   & 0.7   & 0.8   & 0.9   & 0.91  & 0.92  & 0.95  \\ 
\hline
MA (\%)  & 91.45 & 92.96 & 92.06 & 92.65 & 92.63 & 90.33 & 79.97 & 69.13 & 10    \\
ASR (\%) & 97.20 & 96.72 & 96.93 & 96.63 & 96.29 & 96.88 & 96.41 & 97.36 & 100   \\
\bottomrule
\end{tabular}
\begin{tabular}{cc|ccccccccc}
\toprule
\multicolumn{2}{c|}{$\beta\rightarrow$}          & \multirow{2}{*}{0} & \multirow{2}{*}{0.5} & \multirow{2}{*}{0.6} & \multirow{2}{*}{0.7} & \multirow{2}{*}{0.8} & \multirow{2}{*}{0.9} & \multirow{2}{*}{0.91} & \multirow{2}{*}{0.92} & \multirow{2}{*}{0.95}  \\
Defense$\downarrow$   & Metric~(\%)$\downarrow$~ &                    &                      &                      &                      &                      &                      &                       &                       &                        \\
\hline
\multirow{2}{*}{AC}   & TPR~                     & 99.52              & 99.20                & 99.16                & 45.69                & 26.26                & 26.22                & 23.81                 & 13.38                 & 0.00                   \\
                      & FPR~                     & 31.14              & 29.46                & 28.85                & 8.21                 & 7.72                 & 6.21                 & 0.25                  & 7.80                  & 0.00                   \\
\hline
\multirow{2}{*}{SCAn} & TPR~                     & 96.60              & 96.55                & 96.60                & 72.80                & 56.19                & 0.00                 & 0.00                  & 0.00                  & 0.00                   \\
                      & FPR~                     & 0.77               & 1.38                 & 4.60                 & 1.14                 & 0.10                 & 0.00                 & 0.00                  & 0.00                  & 0.00                   \\
\hline
\multirow{2}{*}{Ours} & TPR                      & 99.38              & 99.41                & 98.18                & 97.43                & 97.52                & 94.20                & 24.20                 & 0.00                  & 0.00                   \\
                      & FPR                      & 1.35               & 1.96                 & 1.44                 & 1.15                 & 0.53                 & 1.40                 & 4.17                  & 0.00                  & 0.00                   \\
\bottomrule
\end{tabular}
\vspace{-0.5cm}
\end{table}

Table~\ref{tab:penalty_impact} (bottom subtable) 
shows the TPRs and FPRs of AC, SCAn, and our defense with different penalty factors.
As $\beta$ increased (up to $\beta = 0.9$), the TPR of our defense decreased from $99.38\%$ to $94.20\%$ while FPR was almost unaffected. This shows that the adversary gained a small advantage with $\beta = 0.9$.
On the other hand, the other defenses achieved limited or poor robustness compared to ours with the same $\beta$ values.
With $\beta \geq 0.91$, AC, SCAn, and our method defense failed to counter the attack. However, 
looking at Table~\ref{tab:penalty_impact} (top subtable) we can see the main accuracy degraded with these high $\beta$ values, which made it easy to reject the model due its low performance. 

\vspace{-0.25cm}
\section{Conclusion}
\label{sec_conclusion}
\vspace{-0.25cm}

In this paper, we conducted a layer-wise feature analysis of the behavior of benign and poisoned samples generated by attacked DNNs. We found that the feature difference between benign and poisoned samples tends to reach the maximum at a critical layer, which can be easily located based on the behaviors of benign samples. Based on this finding, we proposed a simple yet effective backdoor detection to determine whether a given suspicious testing sample is poisoned by analyzing the differences between its features and those of a few local benign samples. Our extensive experiments on benchmark datasets confirmed the effectiveness of our detection. We hope our work can provide a deeper understanding of attack mechanisms, to facilitate the design of more effective and efficient backdoor defenses and more secure DNNs.  

\section*{Acknowledgments}
This research was funded by the European Commission (projects H2020-871042 ``SoBigData++'' and H2020-101006879 ``MobiDataLab''), the Government of Catalonia (ICREA Acad\`emia Prize to J.Domingo-Ferrer, grant no. 2021 SGR 00115, and FI\_B00760 grant to N. Jebreel), and MCIN/AEI/ 10.13039/501100011033 and ``ERDF A way of making Europe'' under grant PID2021-123637NB-I00 ``CURLING''. 
The authors are with the UNESCO Chair in Data Privacy, but the views in this paper are their own and are not necessarily shared by UNESCO.


\vspace{-0.5cm}
\bibliographystyle{splncs04}
\bibliography{my_bib}

\appendix
\section{Detailed method}
\label{sec_meth_detailed}

Algorithm~\ref{alg1} summarizes our defense.

\setcounter{algorithm}{1}
\begin{algorithm}[ht!]
    \caption{Detecting backdoor attacks via layer-wise feature analysis }\label{alg1}
    \hspace*{\algorithmicindent} \textbf{Input}: Suspicious trained DNN $f_s$; Validation samples $X_{val}$; Threshold $\tau$; Suspicious input $\bm{x_s}$ \\
    \hspace*{\algorithmicindent} \textbf{Output:} Boolean value (True/False) tells if $\bm{x_s}$ is poisoned. 
    \begin{algorithmic}[1]

    \For{\texttt{each potential target class $t\in\{1, \ldots, \mathcal{C}\}$}}  \Comment{An offline loop conducted for one time only}
        \State $X_{t_{val}}\gets$Split $t$'s benign samples from $X_{val}$
        \State $m\gets|X_{t_{val}}|$
        \State $\{\bm{a_{i}^{\lfloor L/2 \rfloor}}, \ldots, \bm{a_{i}^L}\}_{i = 1}^{m}\gets$Layers' features generated by $f_s$ for $\{\bm{x_{i}}\in X_{t_{val}}\}$
        \State $\bm{\hat{a}_t^l} \gets \frac{1}{m}\sum_{i=1}^{m} \bm{a_{i}^l}$ \Comment{Estimate $t$'s centroid at layer $l\in\{\lfloor L/2\rfloor, \ldots, L\}$}
        \State $cs_{i}^l\gets\Call{CosineSimilarity}{\bm{a_{i}^{l}}, \bm{\hat{a}_t^{l}}}$ \Comment{Similarity of $\bm{a_{i}^{l}}$ to its centroid}
        \State $\hat{cs}_t^l\gets\frac{1}{m}\sum_{i=1}^{m} cs_{i}^l$ \Comment{Aggregate computed benign similarities at layer $l$}
        \State $LOI_t\gets\Call{IdentifyLayerOfInterest}{\{\hat{cs}_t^{\lfloor L/2\rfloor},\ldots, \hat{cs}_t^{L}\}}$ 
        \State $cs_{i}\gets cs_{i}^{LOI_t - 2} + cs_{i}^{LOI_t - 1} + cs_{i}^{LOI_t}$
        \State $\mu_t, \sigma_t\gets \Call{MEAN}{\{cs_{i}\}_{i = 1}^{m}}, \Call{STD}{\{cs_{i}\}_{i = 1}^{m}}$
     \EndFor
     
     \State $IsPoisoned\gets False$
     \State $\hat{y_s}\gets f_s(\bm{x_s})$ \Comment{$\hat{y_s}$ is the predicted class by $f_s$ for $\bm{x_s}$}
     
     \For{\texttt{each potential target class $t\in\{1, \ldots, \mathcal{C}\}$}}
      \If{$\hat{y_s}= t$}
            \State $\{cs_{s}^{LOI_{t} - 2} , cs_{j}^{LOI_{t} - 1} , cs_{j}^{LOI_{t}}\}\gets \{\Call{CosineSimilarity}{\bm{a_{s}^{l}}, \bm{\hat{a}_t^l}}\}_{l = LOI_{t} - 2}^{LOI_{t}}$ 
            \State $cs_{s}\gets cs_{s}^{LOI_{t} - 2} + cs_{s}^{LOI_{t} - 1} + cs_{s}^{LOI_{t}}$
            \If{$cs_{s} < (\mu_{t} - \tau\times\sigma_{t})$}
                \State $IsPoisoned\gets True$
            \EndIf
      \EndIf
     \EndFor
     \State \textbf{return} $IsPoisoned$
    
     \Procedure{IdentifyLayerOfInterest}{$\{\hat{cs}^{\lfloor L/2 \rfloor},\ldots, \hat{cs}^{L}\}$}
       \State $max_{diff}\gets\hat{cs}^{\lfloor L/2 \rfloor + 1} - \hat{cs}^{\lfloor L/2 \rfloor}$
       \State $LOI\gets \lfloor L/2 \rfloor + 1$
        \For{\texttt{$l\in\{\lfloor L/2 \rfloor + 2, \ldots, L\}$}}
            \State $l_{diff}\gets\hat{cs}^{l} - \hat{cs}^{l-1}$
            \If{$l_{diff} > max_{diff}$}
                \State $max_{diff}\gets l_{diff}$
                \State $LOI\gets l$
            \EndIf
         \EndFor
    \State \textbf{return} $LOI$
    \EndProcedure
    \end{algorithmic}
    \end{algorithm}

For each potential target class $t \in \{1, \ldots \mathcal{C}\}$, we first feed the available $m$ benign samples to $f_s$ and extract their intermediate features in the second half of layers to obtain the set $\{(\bm{a_{i}^{L/2}}, \ldots, \bm{a_{i}^L})\}_{i=1}^{m}$ (if $L$ is odd, take the
integer part of $L/2$ instead of $L/2$ here and in what follows). Note that we can reduce computation
by focusing on the second half of layers because 
the LOI and the two layers before the LOI are among 
the latter layers of the DNN.  
After that, we compute the layer-wise centroids of the extracted features for each layer $l\in\{L/2, \ldots, L\}$ (Line 5).
Then, we compute the cosine similarity between the benign features of each layer and their corresponding centroid (Line 6).

Then, we aggregate the computed similarities to approximate the similarity centroid in each layer $l \in \{L/2, (L/2)+1, \ldots, L-1, L\}$ (Line 7).
Next, we use $\{\hat{cs}_t^{L/2}, \hat{cs}_t^{(L/2)+1}, \ldots, \hat{cs}_t^L\}$, to locate the layer of interest $LOI_t$ that contributes most to assigning $t$'s benign samples to their true class $t$ (Lines 20-28). 
We compute the difference between the approximated similarity of each layer and its preceding one, and we identify the layer with the maximum difference as $LOI_t$.  For example, if the maximum difference is $|\hat{cs}_t^{l} - \hat{cs}_t^{l-1}|$, then layer $l$ is the layer of interest.

Once we locate $LOI_t$, we estimate the behavior of benign samples in that layer and in the two layers previous to it. 
For each sample $\bm{x_{i}}\in X_{t_{val}}$, we sum up its computed cosine similarities in the three layers (Line 9).
After computing the summed similarities of the $m$ samples and obtaining the set $\{cs_{i}\}_{i = 1}^{m}$, we compute the mean $\mu_t$ and the standard deviation $\sigma_t$ of the set.

To detect potentially poisoned samples, for any suspicious incoming input $\bm{x_s}$ classified as $t$ by $f_s$ at inference time, we extract its features in $LOI_t$ and the two preceding layers, compute their cosine similarities to the corresponding estimated centroids $\{cs_{s}^{LOI_t - 2}, cs_{s}^{LOI_t - 1}, cs_{s}^{LOI_t}\}$, and sum them up to get $cs_s$.
Then, we identify $\bm{x_s}$ as a potentially poisoned sample if $cs_s < \mu_t - \tau\times\sigma_t$, where $\tau$ is an input threshold chosen by the defender that provides a reasonable trade-off between  the true positive rate TPR and the false positive rate FPR.
Figure~\ref{fig:lc_hist_distribution} shows an example of the distributions of the summed cosine similarities of benign and poisoned features to the estimated benign centroids (in the three identified layers) under the label-consistent attack of \cite{turner2019label}. 

\setcounter{figure}{5}
\begin{figure*}[!ht]
    \centering
    \begin{subfigure}{0.43\textwidth}
      \centering
      \includegraphics[width=1\linewidth]{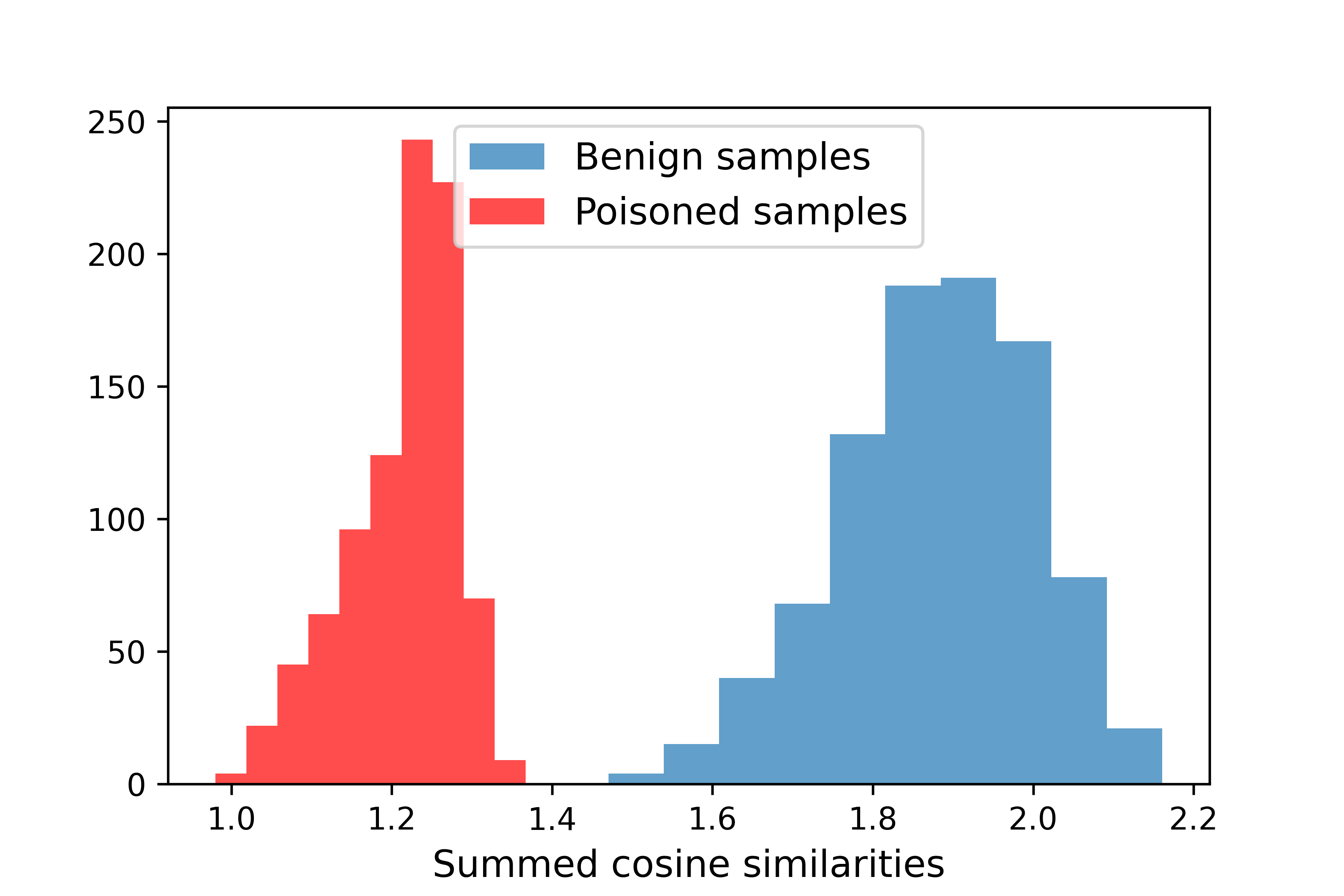}
      \caption{CIFAR10-ResNet18}
    \end{subfigure}%
    \hspace{1em}
    \begin{subfigure}{0.43\textwidth}
      \centering
      \includegraphics[width=1\linewidth]{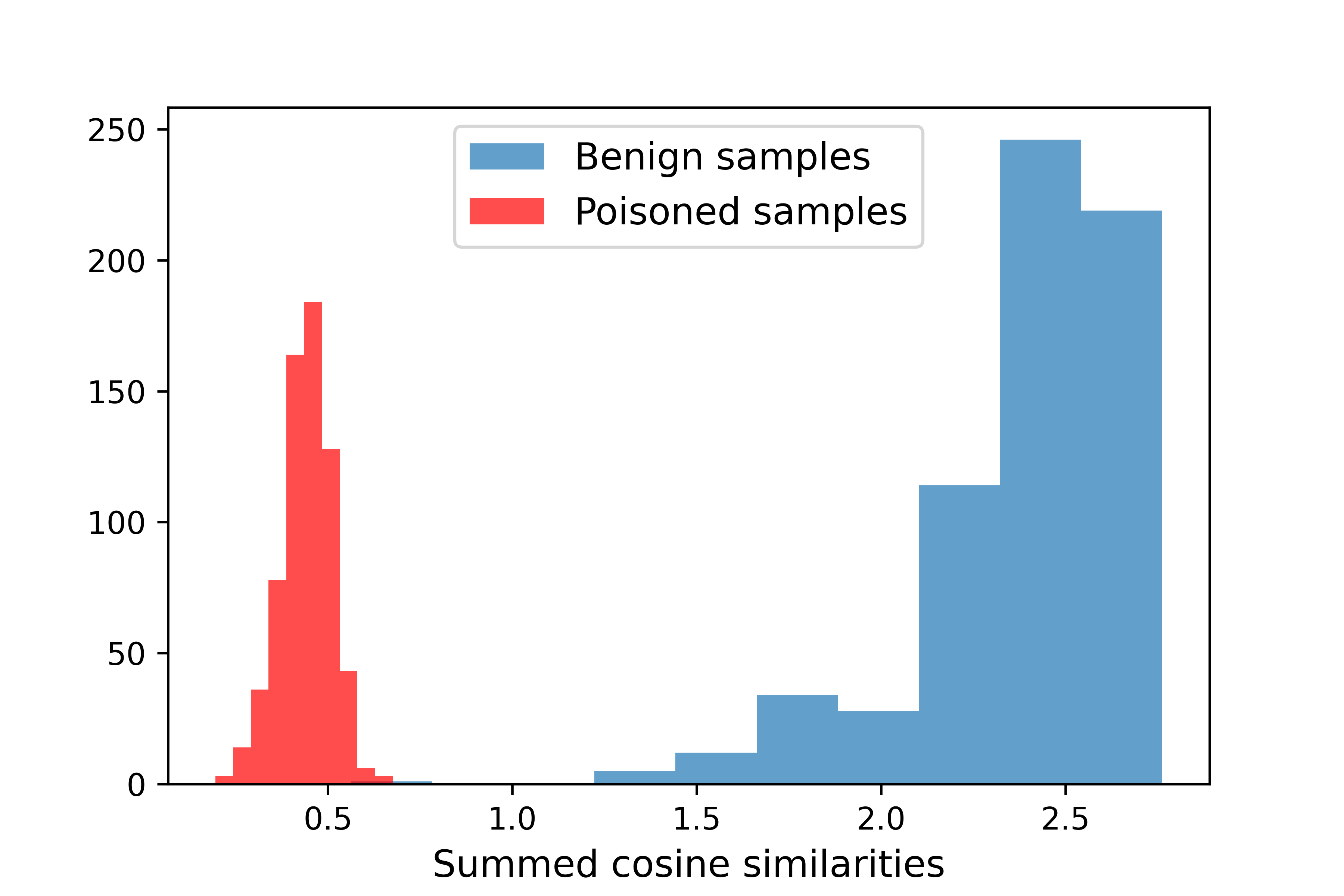}
      \caption{GTSRB-MobileNetV2}
    \end{subfigure}%
\caption{Distributions of the summed cosine similarities of benign and poisoned samples under the label-consistent attack on CIFAR10 with ResNet18 and GTSRB with MobileNetV2 benchmarks}
\label{fig:lc_hist_distribution}
\end{figure*}

\section{Additional Results on Layer-wise Feature Analysis}
\label{sec_more_analysis}
\begin{figure}[t!]
    \centering
    \begin{subfigure}{0.5\textwidth}
      \centering
      \includegraphics[width=1\linewidth]{cifar10_resnet18_badnets_layer_sim.png}
      \caption{BadNets}
    \end{subfigure}%
    \begin{subfigure}{0.5\textwidth}
      \centering
      \includegraphics[width=1\linewidth]{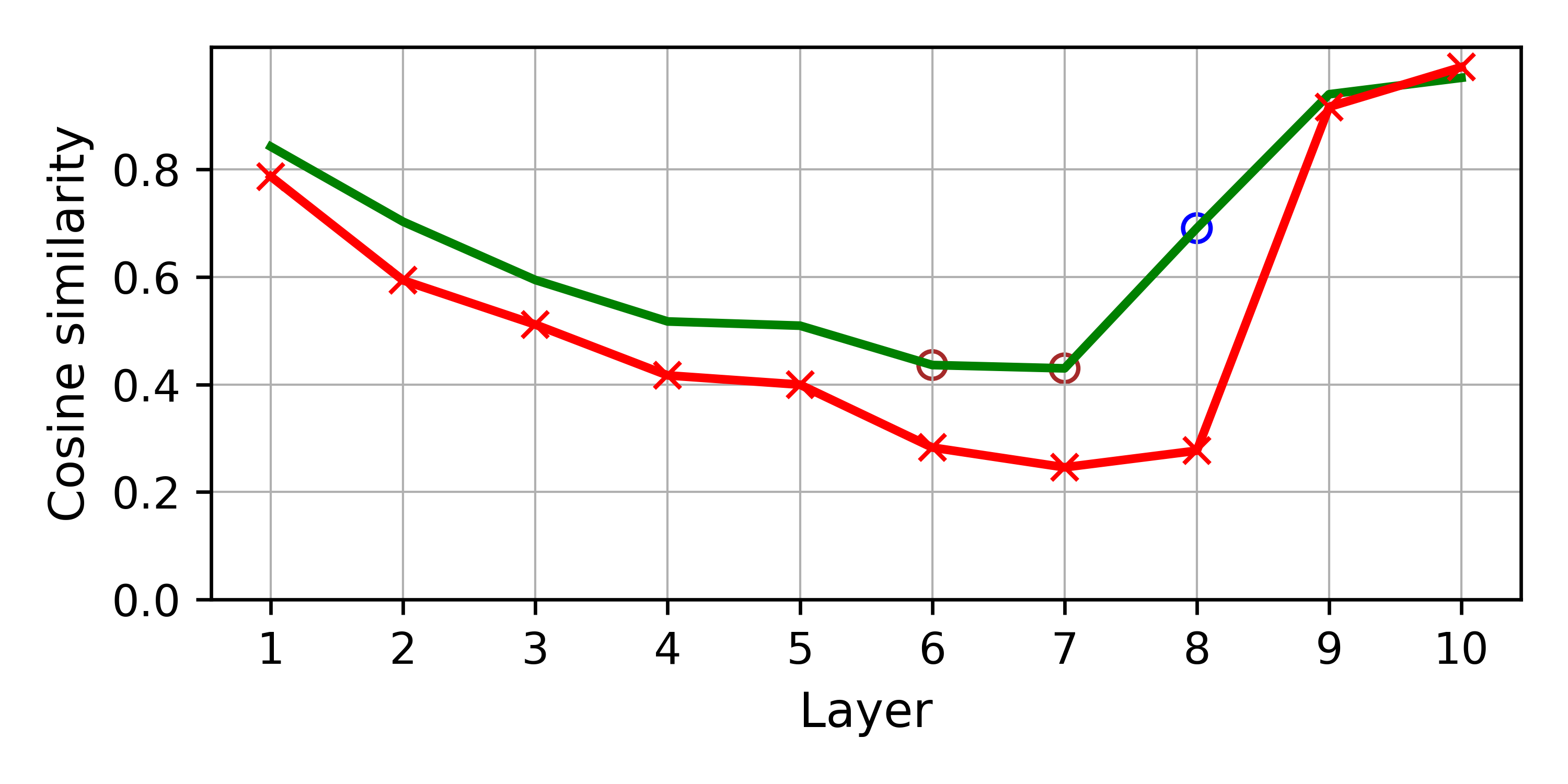}
      \caption{Blended}
    \end{subfigure}%
    
     \begin{subfigure}{0.5\textwidth}
      \centering
      \includegraphics[width=1\linewidth]{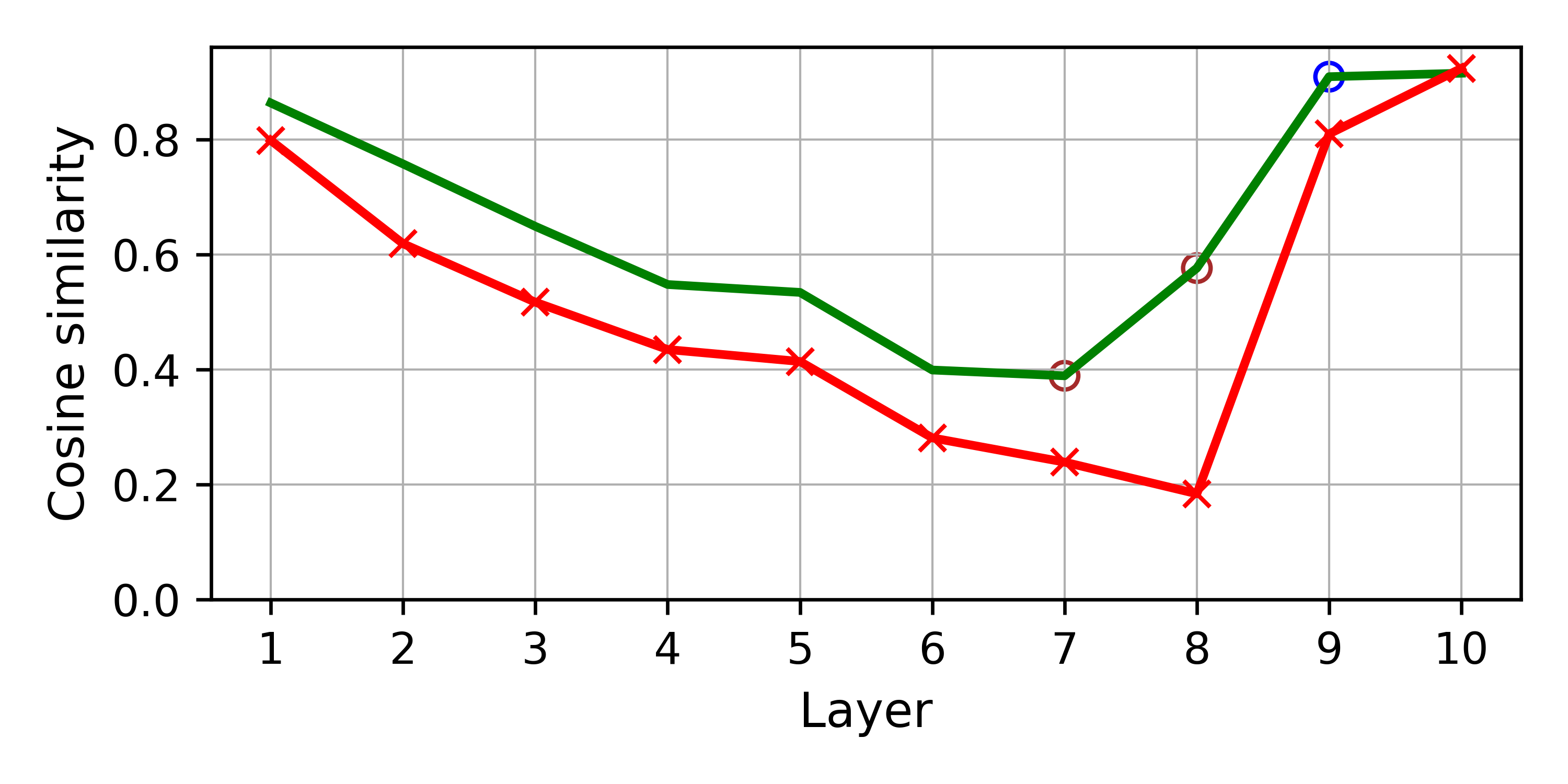}
      \caption{Label-consistent}
    \end{subfigure}%
     \begin{subfigure}{0.5\textwidth}
      \centering
      \includegraphics[width=1\linewidth]{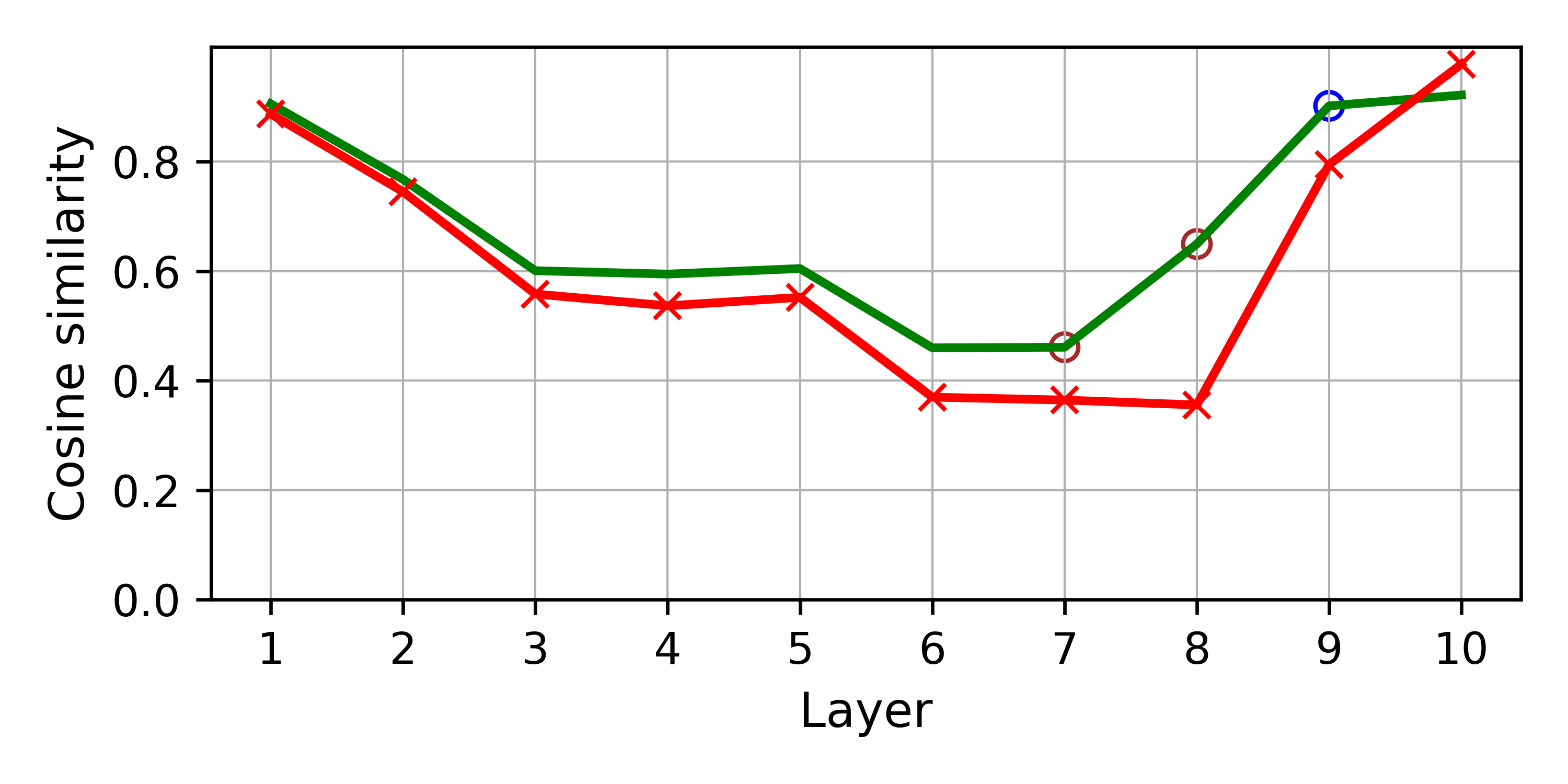}
      \caption{WaNet}
    \end{subfigure}%
    
     \begin{subfigure}{0.5\textwidth}
      \centering
      \includegraphics[width=1\linewidth]{cifar10_resnet18_issba_layer_sim.png}
      \caption{ISSBA}
    \end{subfigure}%
     \begin{subfigure}{0.5\textwidth}
      \centering
      \includegraphics[width=1\linewidth]{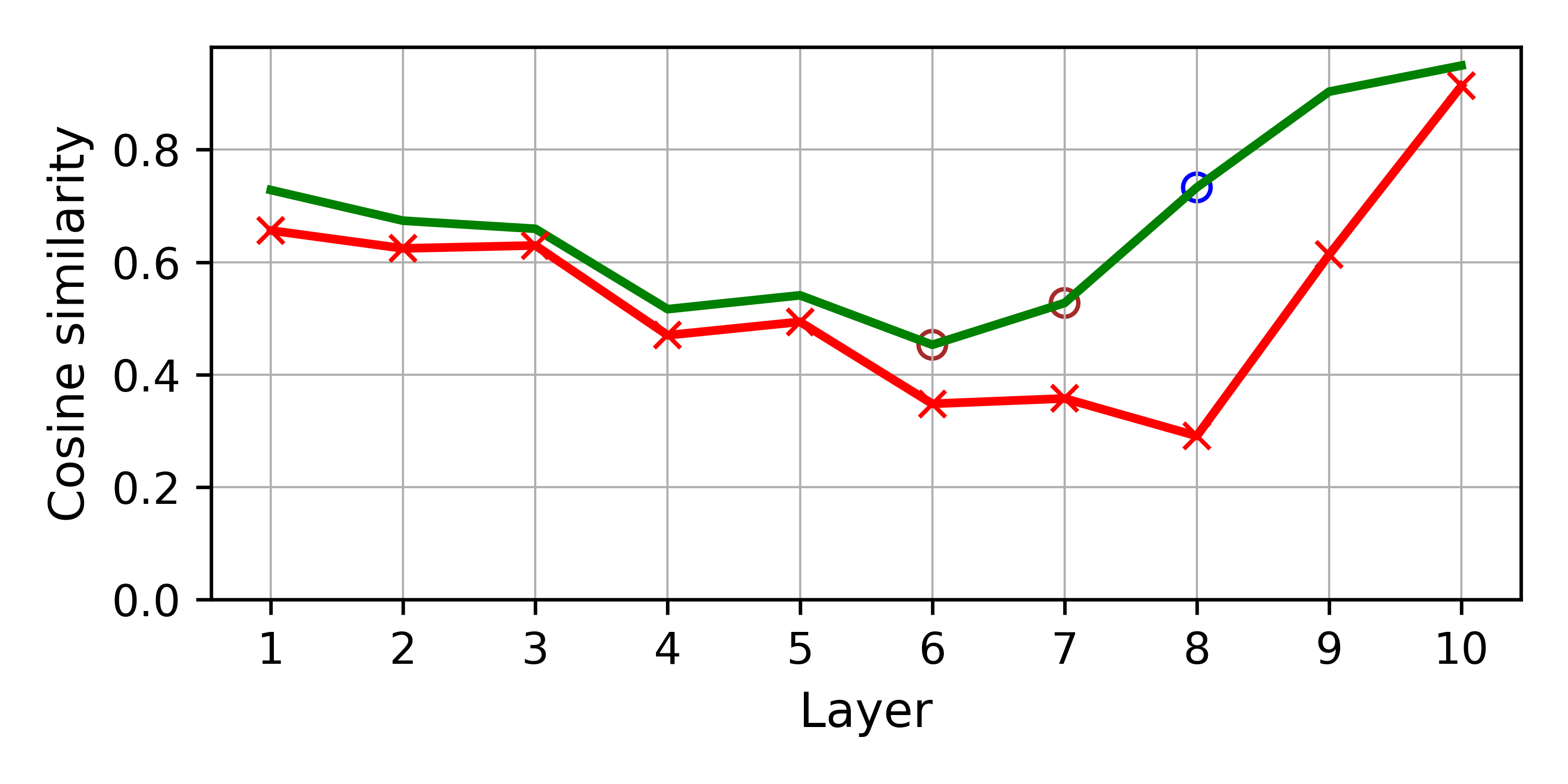}
      \caption{IAD}
    \end{subfigure}
    
\caption{Layer-wise behavior of benign and poisoned samples w.r.t. the target class in the CIFAR10-ResNet18 benchmark, under all implemented attacks}
\label{fig:cifar10_resnet18_analysis}
\end{figure}

\begin{figure}[t!]
    \centering
    \begin{subfigure}{0.5\textwidth}
      \centering
      \includegraphics[width=1\linewidth]{gtsrb_mobilenetv2_badnets_layer_sim.png}
      \caption{BadNets}
    \end{subfigure}%
    \begin{subfigure}{0.5\textwidth}
      \centering
      \includegraphics[width=1\linewidth]{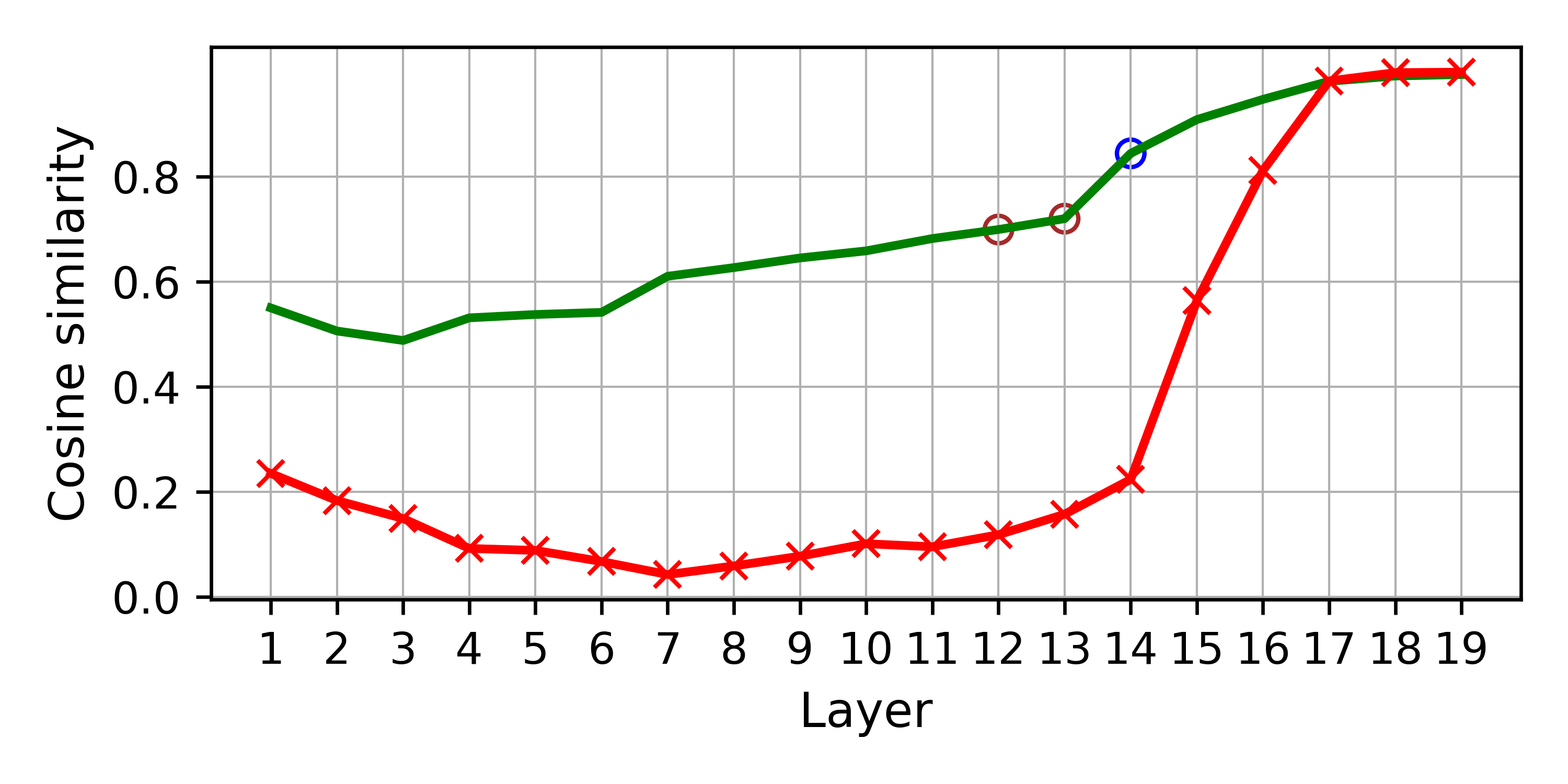}
      \caption{Blended}
    \end{subfigure}%
    
     \begin{subfigure}{0.5\textwidth}
      \centering
      \includegraphics[width=1\linewidth]{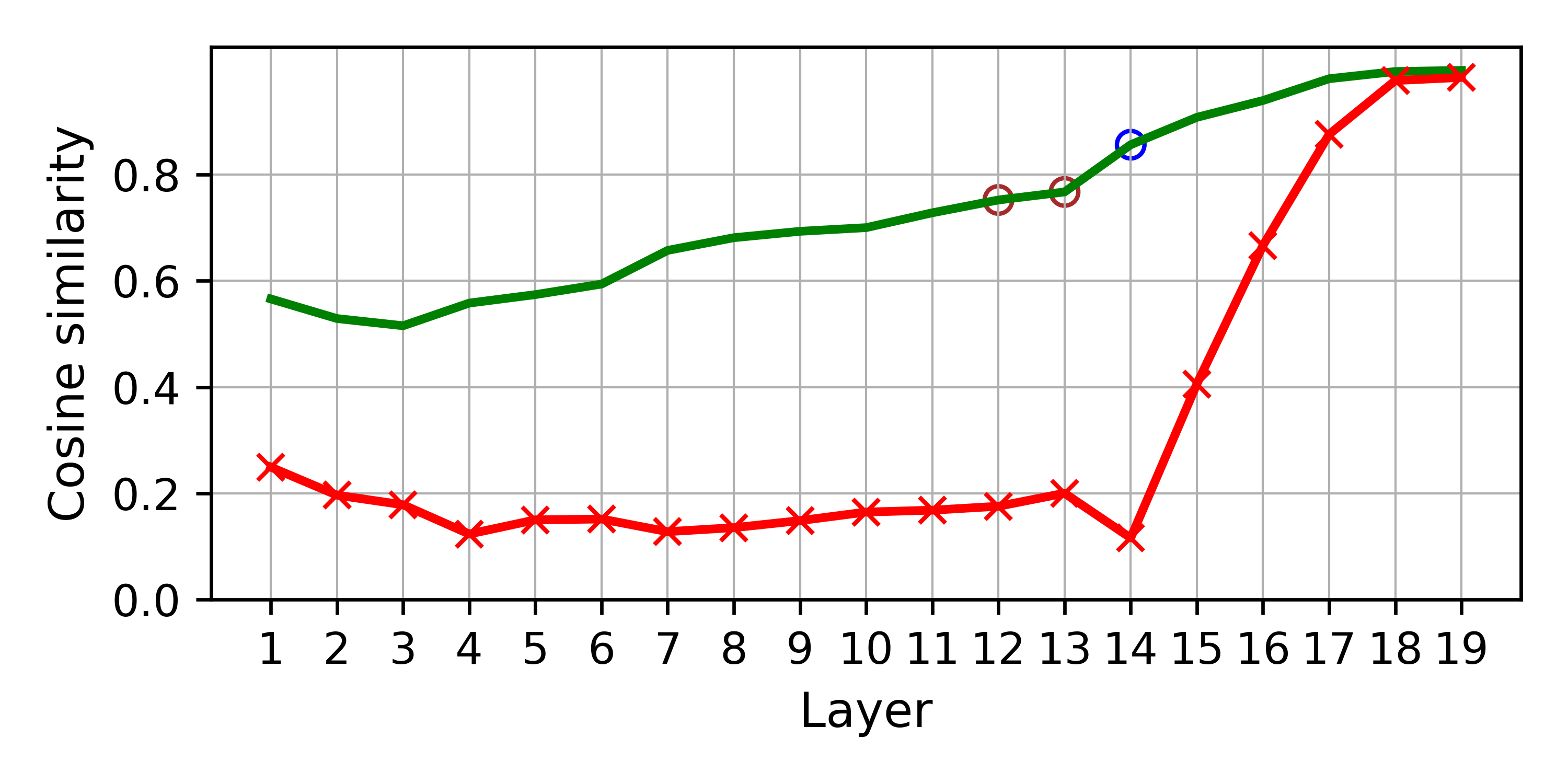}
      \caption{Label-consistent}
    \end{subfigure}%
     \begin{subfigure}{0.5\textwidth}
      \centering
      \includegraphics[width=1\linewidth]{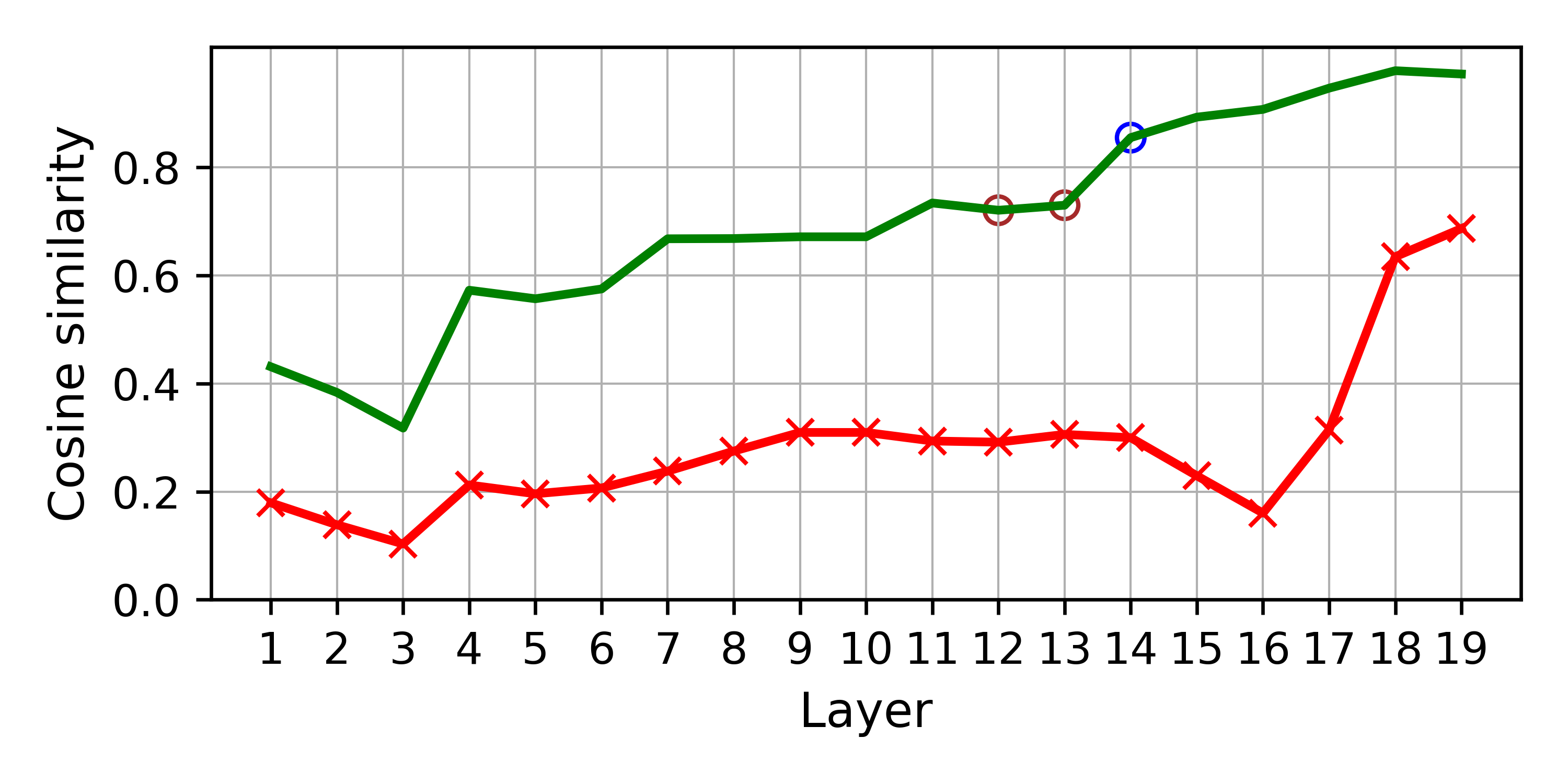}
      \caption{WaNet}
    \end{subfigure}%
    
     \begin{subfigure}{0.5\textwidth}
      \centering
      \includegraphics[width=1\linewidth]{gtsrb_mobilenetv2_issba_layer_sim.png}
      \caption{ISSBA}
    \end{subfigure}%
     \begin{subfigure}{0.5\textwidth}
      \centering
      \includegraphics[width=1\linewidth]{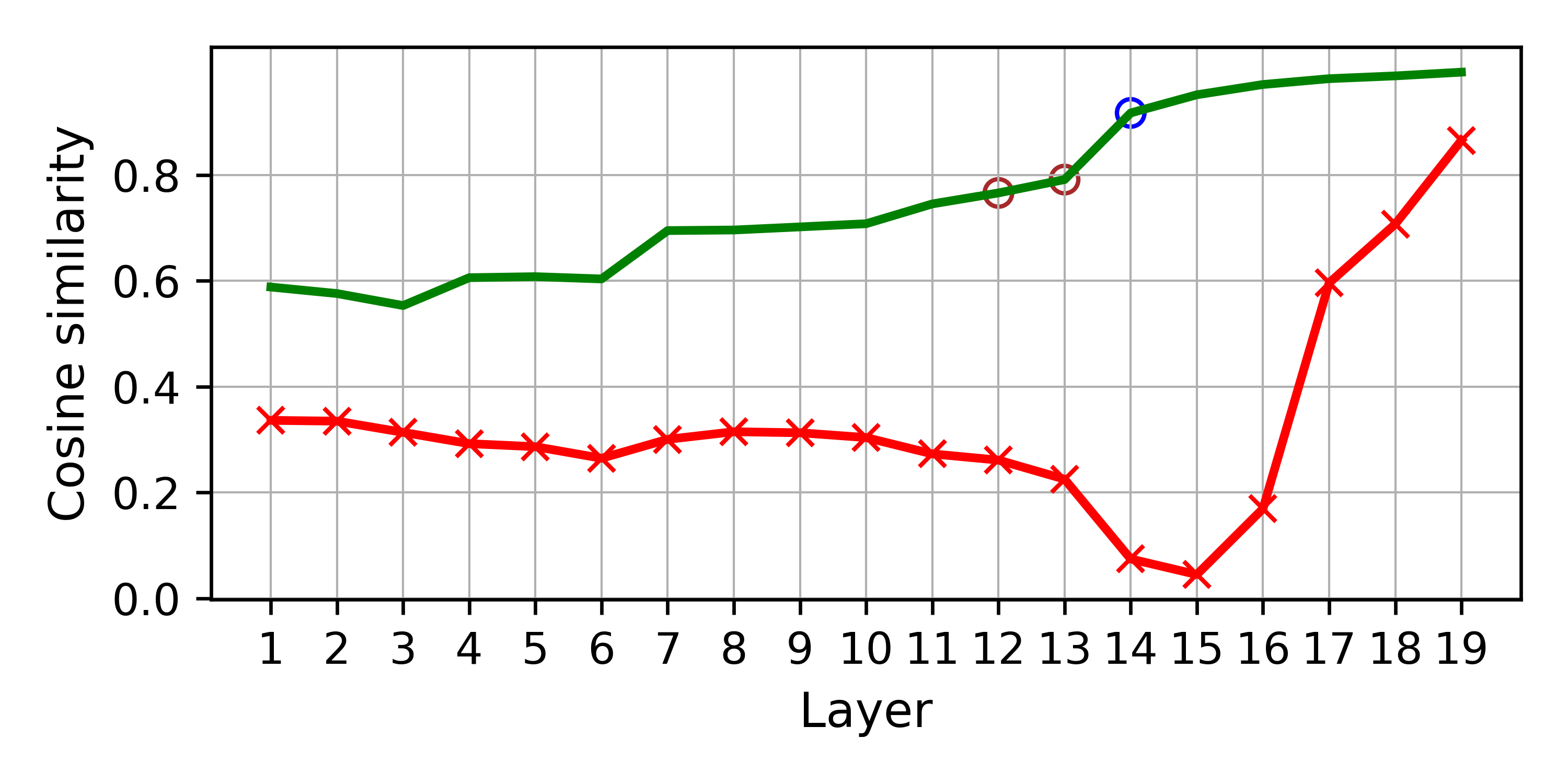}
      \caption{IAD}
    \end{subfigure}%
\caption{Layer-wise behavior of benign and poisoned samples w.r.t. the target class in the GTSRB-MobileNetV2 benchmark, under all implemented attacks}
\label{fig:gtsrb_mobilenetv2_analysis}
\end{figure}

Figure~\ref{fig:cifar10_resnet18_analysis} shows the layer-wise behavior of benign and poisoned features w.r.t. the target class on the CIFAR10-ResNet18 benchmark under all the used attacks. 
Figure~\ref{fig:gtsrb_mobilenetv2_analysis} shows the same on the GTSRB-MobileNetV2 benchmark.

It can be seen that the layer with the maximum difference in cosine similarity is likely to be one of the three circled layers (the LOI and the two preceding layers). This happens in all cases, except for WaNet on GTSRB-MobileNetV2.
We can also notice that the layer-wise gaps are smaller for WaNet, which is stealthier than the other attacks. 
Nevertheless, no matter how stealthy the attack is, the difference is always evident in one of the circled layers.

\section{Additional Discussion}
\label{sec_ablation_studies}

\subsection{Stability Comparison}
\label{sec_stability}

\begin{figure}[t!]
    \centering
    \begin{subfigure}{0.5\textwidth}
      \centering
      \includegraphics[width=1\linewidth]{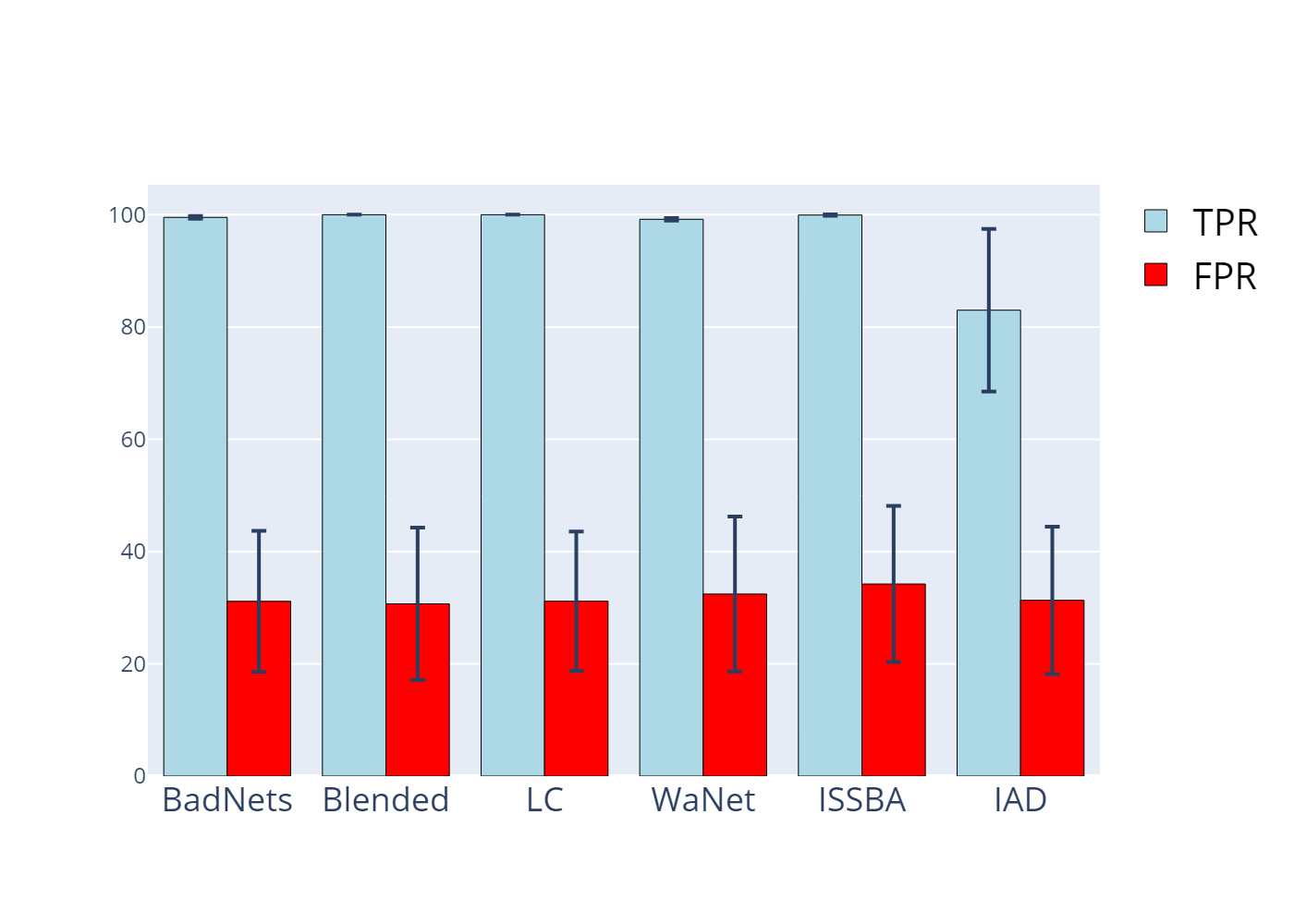}
      \caption{AC}
    \end{subfigure}%
    \begin{subfigure}{0.5\textwidth}
      \centering
      \includegraphics[width=1\linewidth]{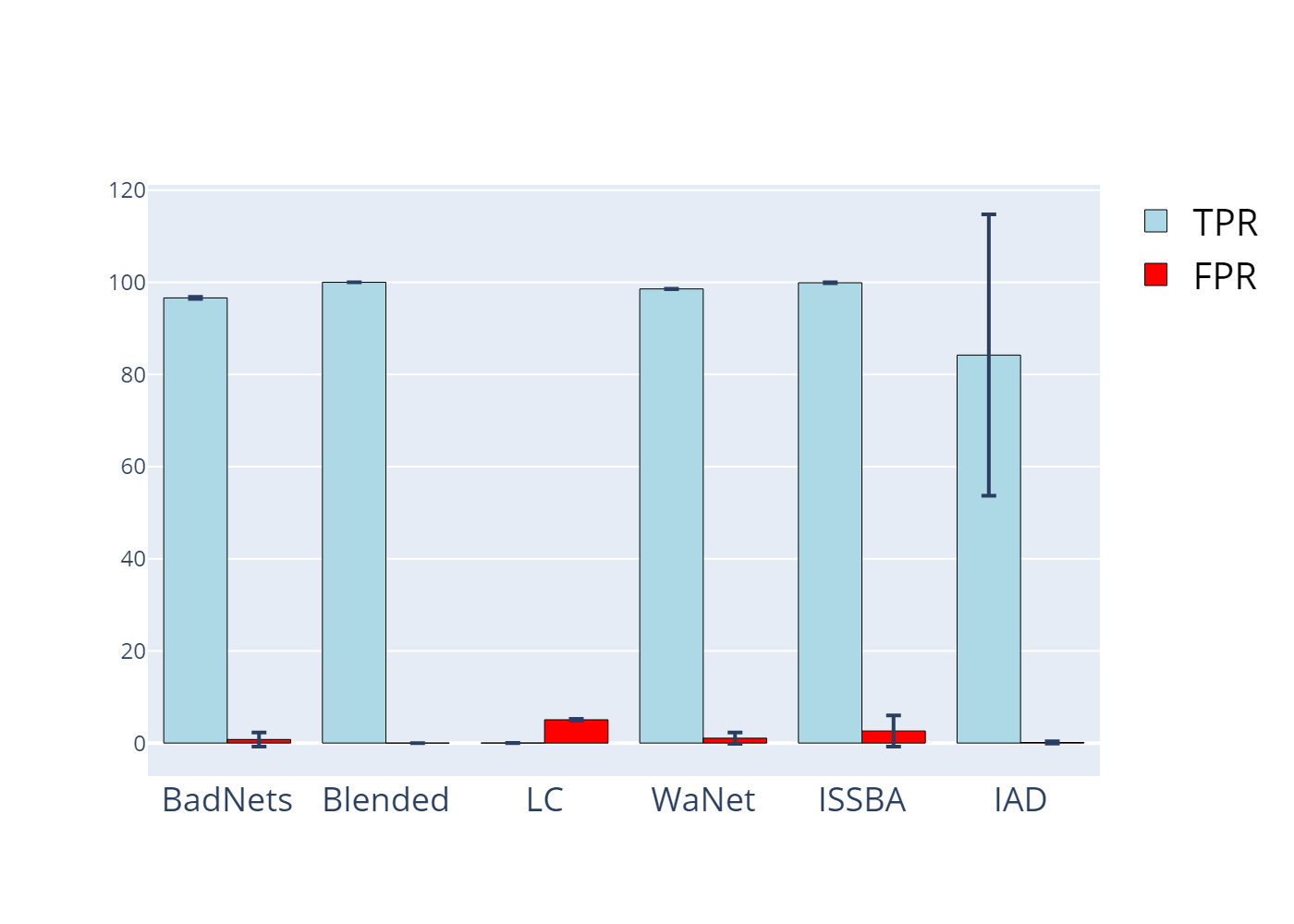}
      \caption{SCAn}
    \end{subfigure}%
    
     \begin{subfigure}{0.5\textwidth}
      \centering
      \includegraphics[width=1\linewidth]{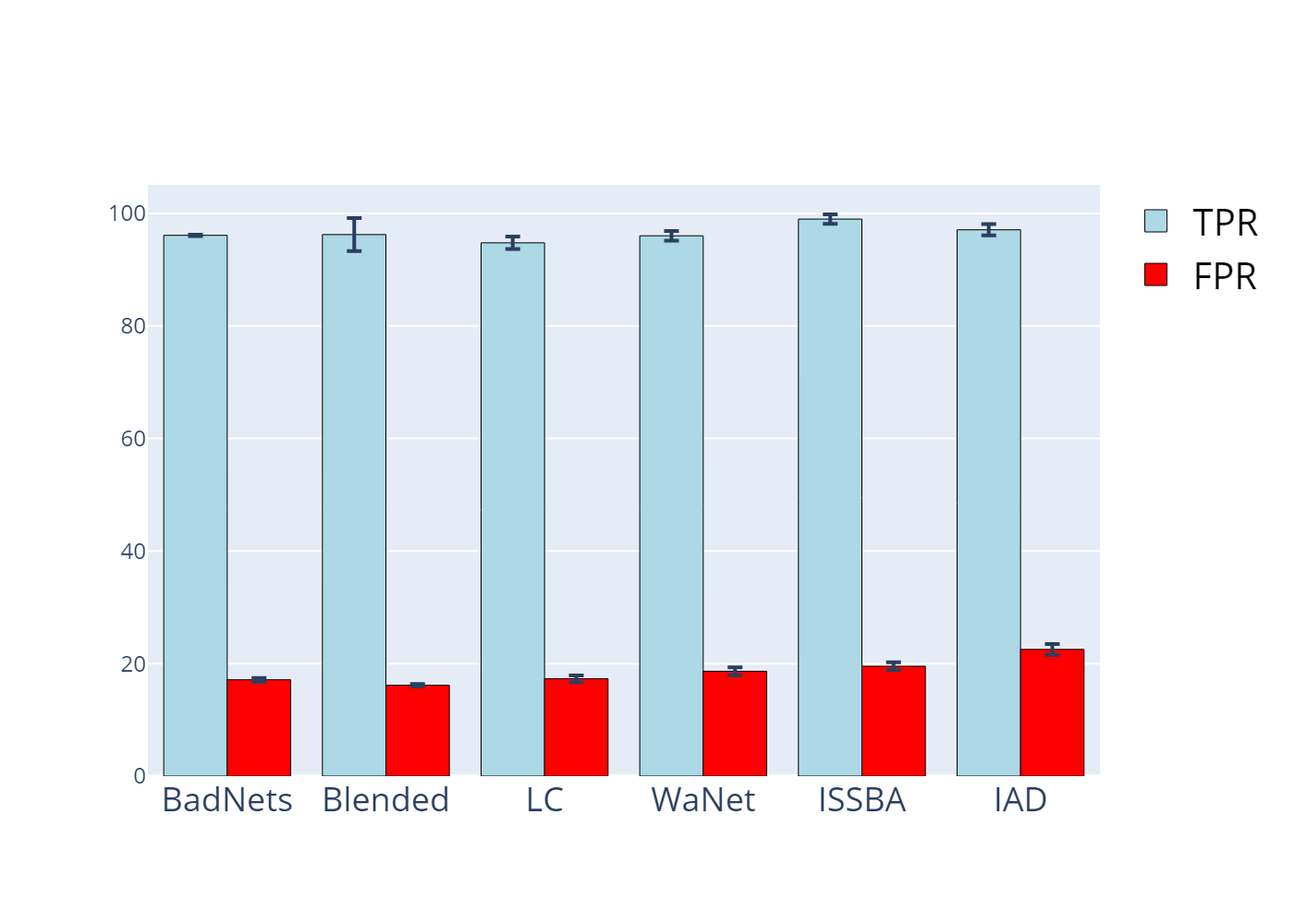}
      \caption{FP}
    \end{subfigure}%
     \begin{subfigure}{0.5\textwidth}
      \centering
      \includegraphics[width=1\linewidth]{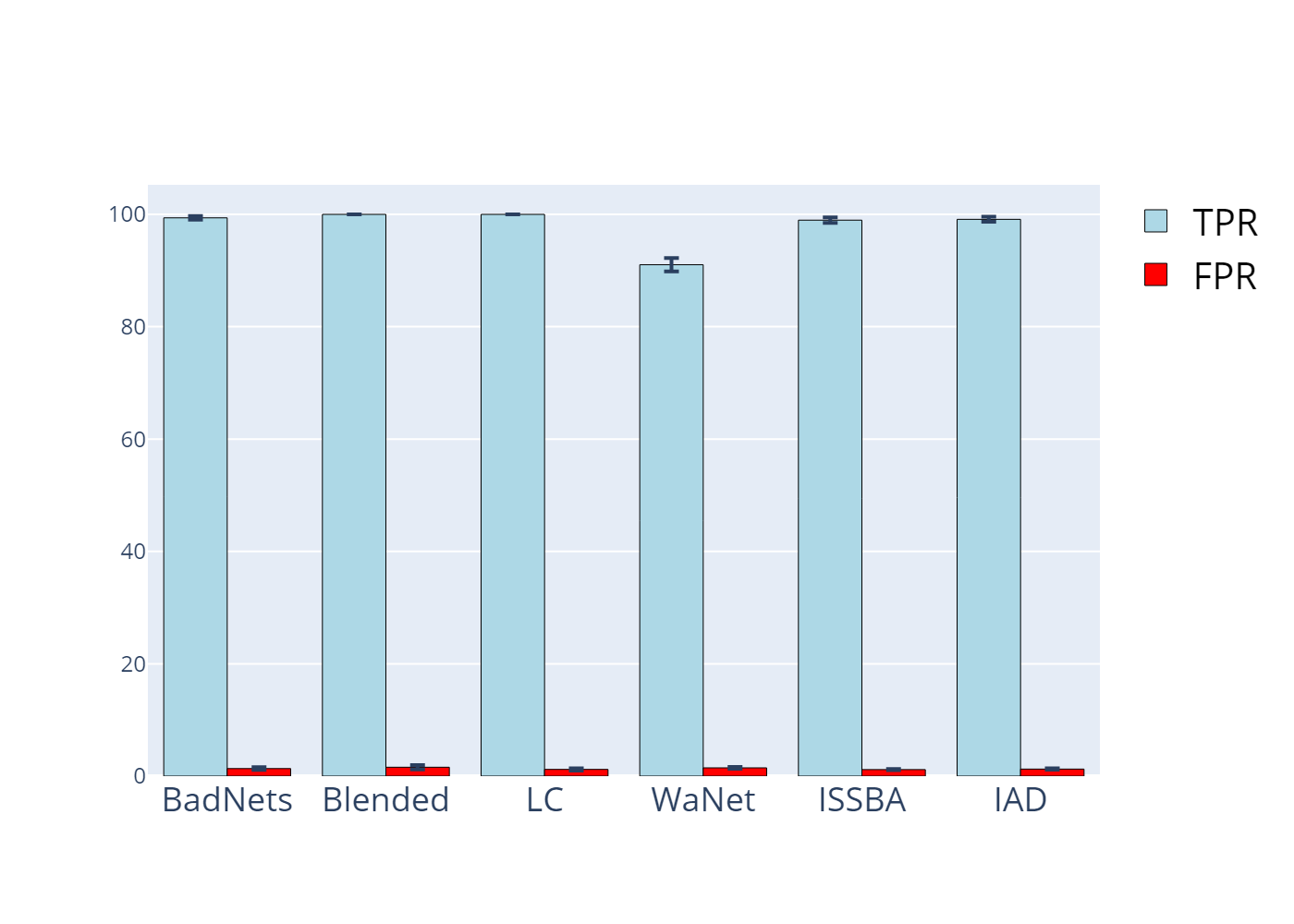}
      \caption{Ours}
    \end{subfigure}%
     
\caption{Stability on the CIFAR10-ResNet18 benchmark}
\label{fig:cifar10_resnet18_stability}
\end{figure}

\begin{figure}[t!]
    \centering
    \begin{subfigure}{0.5\textwidth}
      \centering
      \includegraphics[width=1\linewidth]{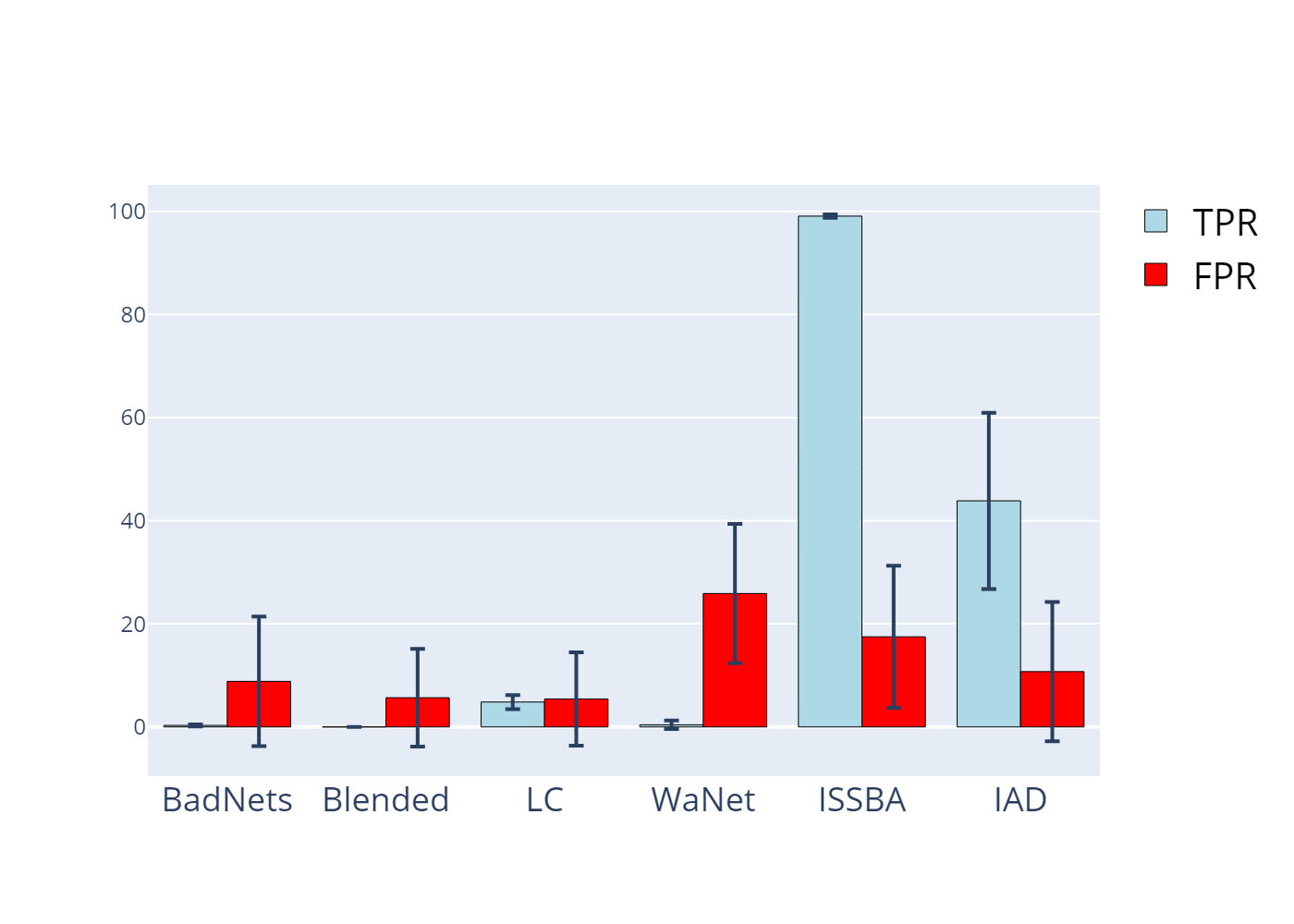}
      \caption{AC}
    \end{subfigure}%
    \begin{subfigure}{0.5\textwidth}
      \centering
      \includegraphics[width=1\linewidth]{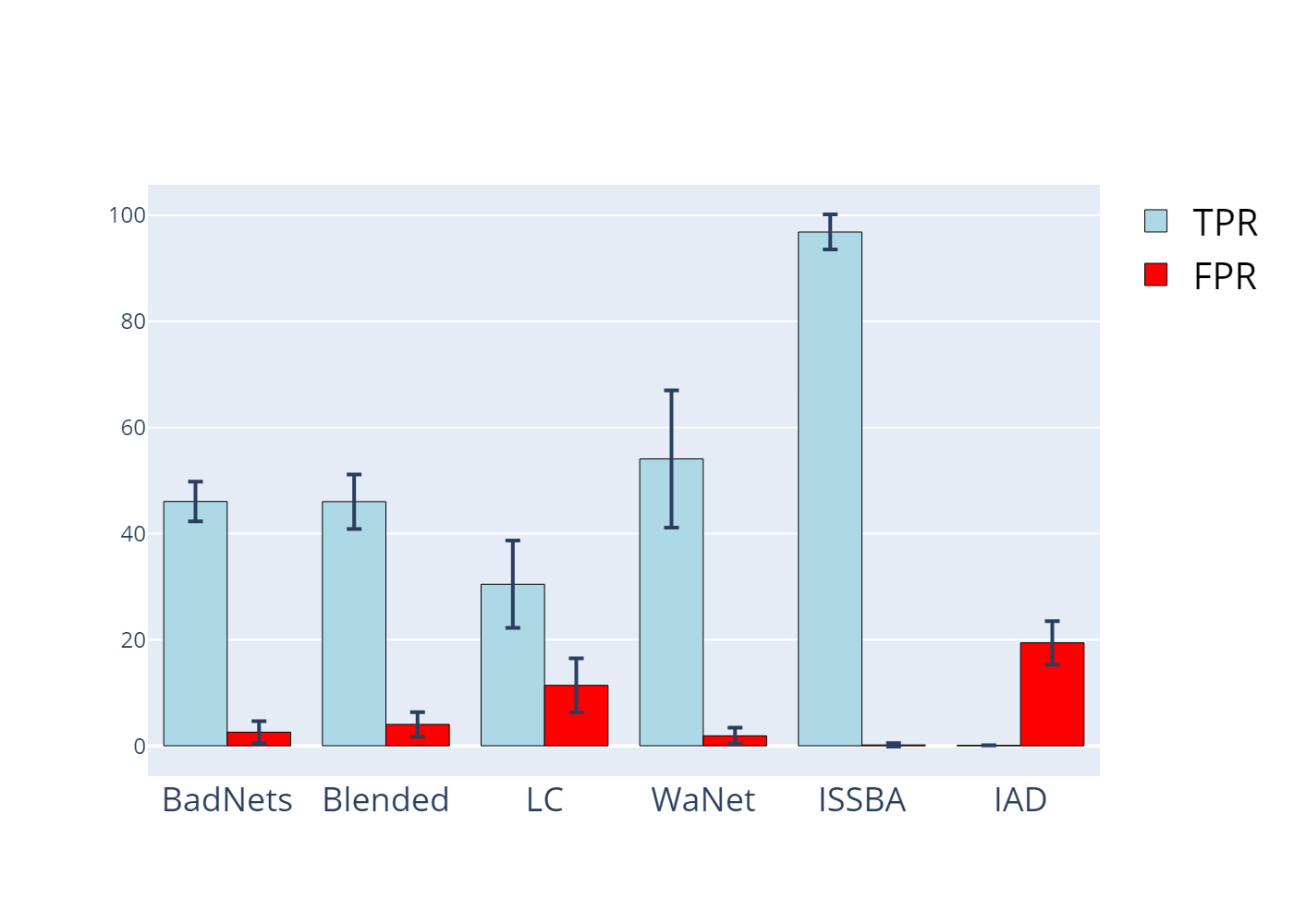}
      \caption{SCAn}
    \end{subfigure}%
    
     \begin{subfigure}{0.5\textwidth}
      \centering
      \includegraphics[width=1\linewidth]{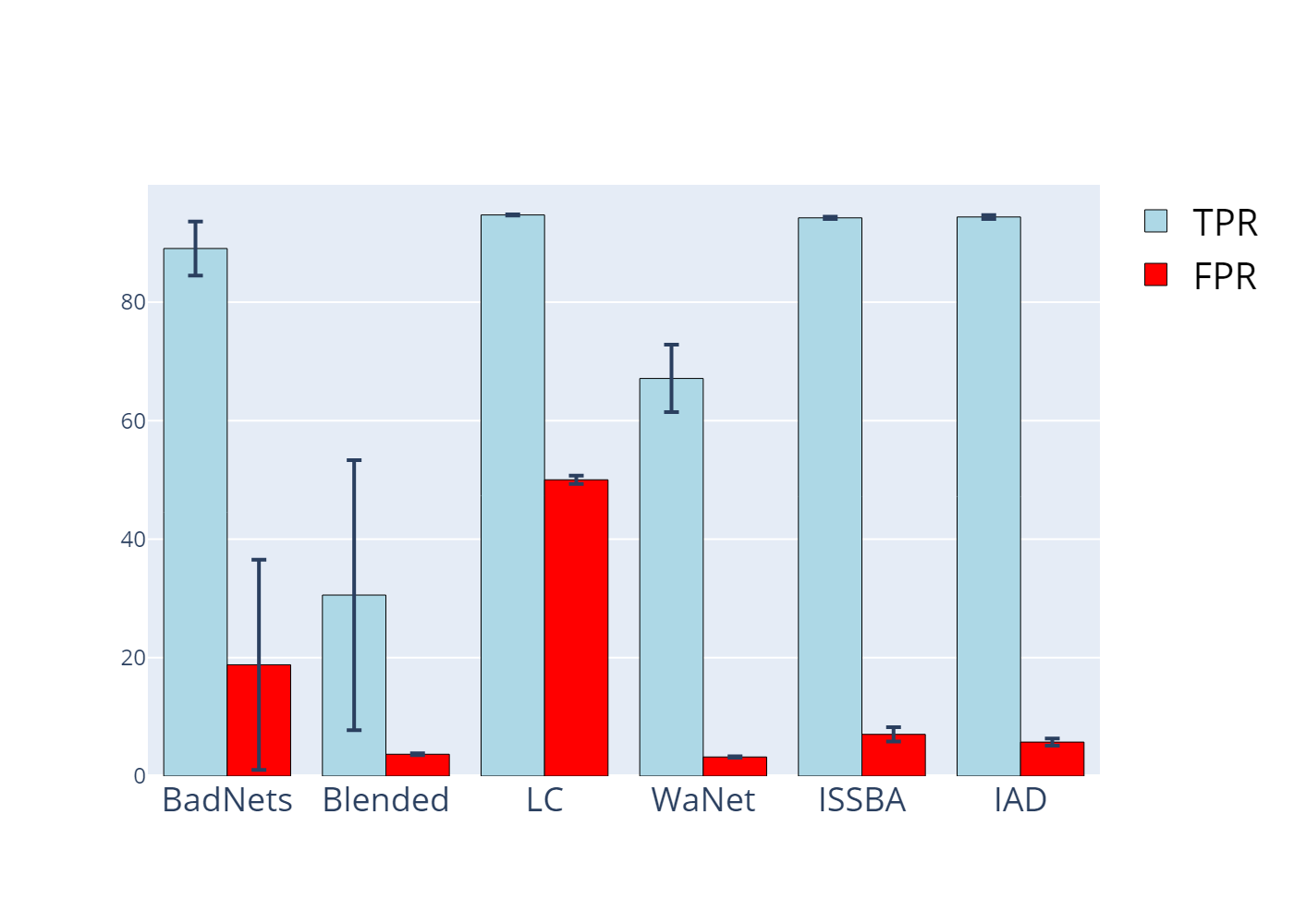}
      \caption{FP}
    \end{subfigure}%
     \begin{subfigure}{0.5\textwidth}
      \centering
      \includegraphics[width=1\linewidth]{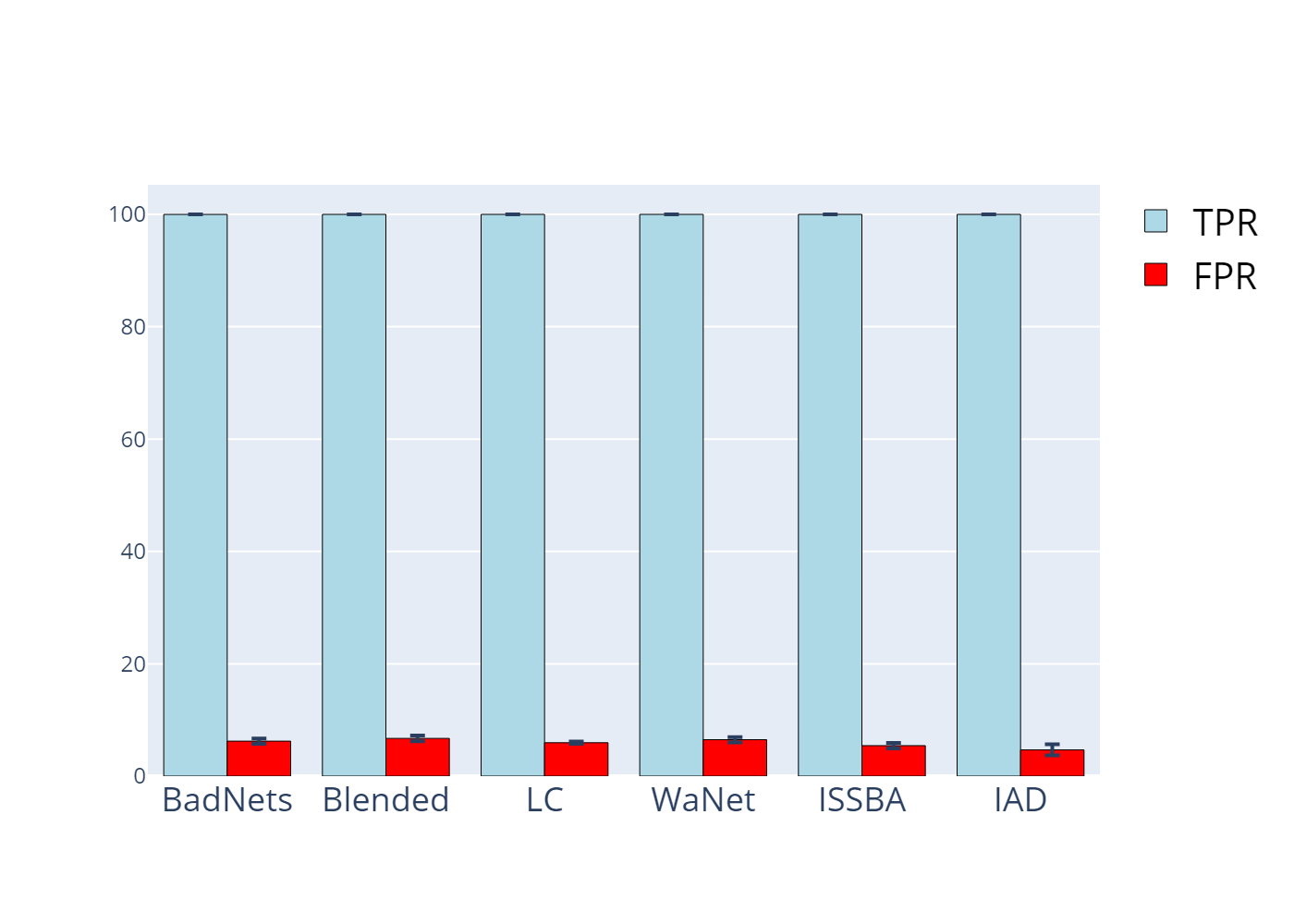}
      \caption{Ours}
    \end{subfigure}%
     
\caption{Stability on the GTSRB-MobileNetV2 benchmark}
\label{fig:gtsrb_mobilenetv2_stability}
\end{figure}

We compared the stability of our defense with that of AC, SCAn, and FP on the CIFAR10-ResNet18 and GTSRB-MobileNetV2 benchmarks.
We ran each defense five times and we report the average TPR and FPR with their standard deviations.
Error bars in Figure~\ref{fig:cifar10_resnet18_stability} and Figure~\ref{fig:gtsrb_mobilenetv2_stability} show that our defense, in general, is more stable than the others.

\subsection{Effectiveness of Cosine Similarity}
\label{sec_cs_effectivness}
We also tried the Euclidean distance as a metric to differentiate between benign and poisoned samples, as we did with cosine similarity.
The only difference was considering any suspicious input with a summed distance greater than the mean of benign samples with $\tau$ standard deviations as potentially poisoned. 
Table~\ref{tab:cs_vs_eucdist} shows the detection performance of our defense with each of the two metrics in the CIFAR10-ResNet18 benchmark under the IAD backdoor attack with different thresholds.
It can be seen that cosine similarity provides a better differentiation between benign and poisoned samples.
A possible explanation is that the direction of features is more important for detection than their magnitude.

\begin{table}[t!]
\centering
\caption{Comparison between Euclidean distance and cosine similarity as metrics to differentiate between benign and poisoned samples ($\pm$: standard deviation). Best scores are in bold.}
\label{tab:cs_vs_eucdist}
\resizebox{\textwidth}{!}{%
\begin{tabular}{c|cccccccccccc} 
\toprule
Threshold$\rightarrow$                                                                                                      & \multicolumn{2}{c}{0.5}                                                                                                                                        & \multicolumn{2}{c}{1}                                                                                                                                & \multicolumn{2}{c}{1.5}                                                                                                                                      & \multicolumn{2}{c}{2}                                                                                                                                        & \multicolumn{2}{c}{2.5}                                                                                                                                      & \multicolumn{2}{c}{3}                                                                                                                                \\ 
\hline
\multicolumn{1}{l|}{\begin{tabular}[c]{@{}l@{}}Evaluation metric$\rightarrow$ \\Similarity metric$\downarrow$\end{tabular}} & TPR\%                                                                          & FPR\%                                                                         & TPR\%                                                                         & FPR\%                                                                & TPR\%                                                                         & FPR\%                                                                        & TPR\%                                                                         & FPR\%                                                                        & TPR\%                                                                         & FPR\%                                                                        & TPR\%                                                                         & FPR\%                                                                \\ 
\hline
Euclidean distance                                                                                                          & \begin{tabular}[c]{@{}c@{}}99.98\\ ($\pm$0.01)\end{tabular}                    & \begin{tabular}[c]{@{}c@{}}\textbf{24.77}\\ ($\pm$\textbf{0.64})\end{tabular} & \begin{tabular}[c]{@{}c@{}}97.12\\ ($\pm$3.13)\end{tabular}                   & \begin{tabular}[c]{@{}c@{}}\textbf{13.64}\\ ($\pm$0.90)\end{tabular} & \begin{tabular}[c]{@{}c@{}}95.73\\ ($\pm$2.41)\end{tabular}                   & \begin{tabular}[c]{@{}c@{}}\textbf{8.67}\\ ($\pm$\textbf{0.36})\end{tabular} & \begin{tabular}[c]{@{}c@{}}70.84\\ ($\pm$10.00)\end{tabular}                  & \begin{tabular}[c]{@{}c@{}}4.69\\ ($\pm$0.36)\end{tabular}                   & \begin{tabular}[c]{@{}c@{}}53.13\\ ($\pm$12.94)\end{tabular}                  & \begin{tabular}[c]{@{}c@{}}3.21\\ ($\pm$1.78)\end{tabular}                   & \begin{tabular}[c]{@{}c@{}}15.71\\ ($\pm$10.61)\end{tabular}                  & \begin{tabular}[c]{@{}c@{}}1.56\\ ($\pm$\textbf{0.06})\end{tabular}  \\
Cosine similarity                                                                                                           & \begin{tabular}[c]{@{}c@{}}\textbf{100.00}\\ ($\pm$\textbf{0.00})\end{tabular} & \begin{tabular}[c]{@{}c@{}}33.65\\ ($\pm$1.11)\end{tabular}                   & \begin{tabular}[c]{@{}c@{}}\textbf{99.99}\\ ($\pm$\textbf{0.00})\end{tabular} & \begin{tabular}[c]{@{}c@{}}18.74\\ ($\pm$\textbf{0.73})\end{tabular} & \begin{tabular}[c]{@{}c@{}}\textbf{99.91}\\ ($\pm$\textbf{0.05})\end{tabular} & \begin{tabular}[c]{@{}c@{}}8.71\\ ($\pm$0.88)\end{tabular}                   & \begin{tabular}[c]{@{}c@{}}\textbf{99.76}\\ ($\pm$\textbf{0.04})\end{tabular} & \begin{tabular}[c]{@{}c@{}}\textbf{3.89}\\ ($\pm$\textbf{0.32})\end{tabular} & \begin{tabular}[c]{@{}c@{}}\textbf{99.12}\\ ($\pm$\textbf{0.45})\end{tabular} & \begin{tabular}[c]{@{}c@{}}\textbf{0.17}\\ ($\pm$\textbf{0.13})\end{tabular} & \begin{tabular}[c]{@{}c@{}}\textbf{95.99}\\ ($\pm$\textbf{3.04})\end{tabular} & \begin{tabular}[c]{@{}c@{}}\textbf{0.40}\\ ($\pm$0.18)\end{tabular}  \\
\bottomrule
\end{tabular}}
\end{table}

\subsection{Runtime Comparison}
\label{sec_runtime}
\begin{figure*}[t!]
    \centering
      \includegraphics[width=0.7\linewidth]{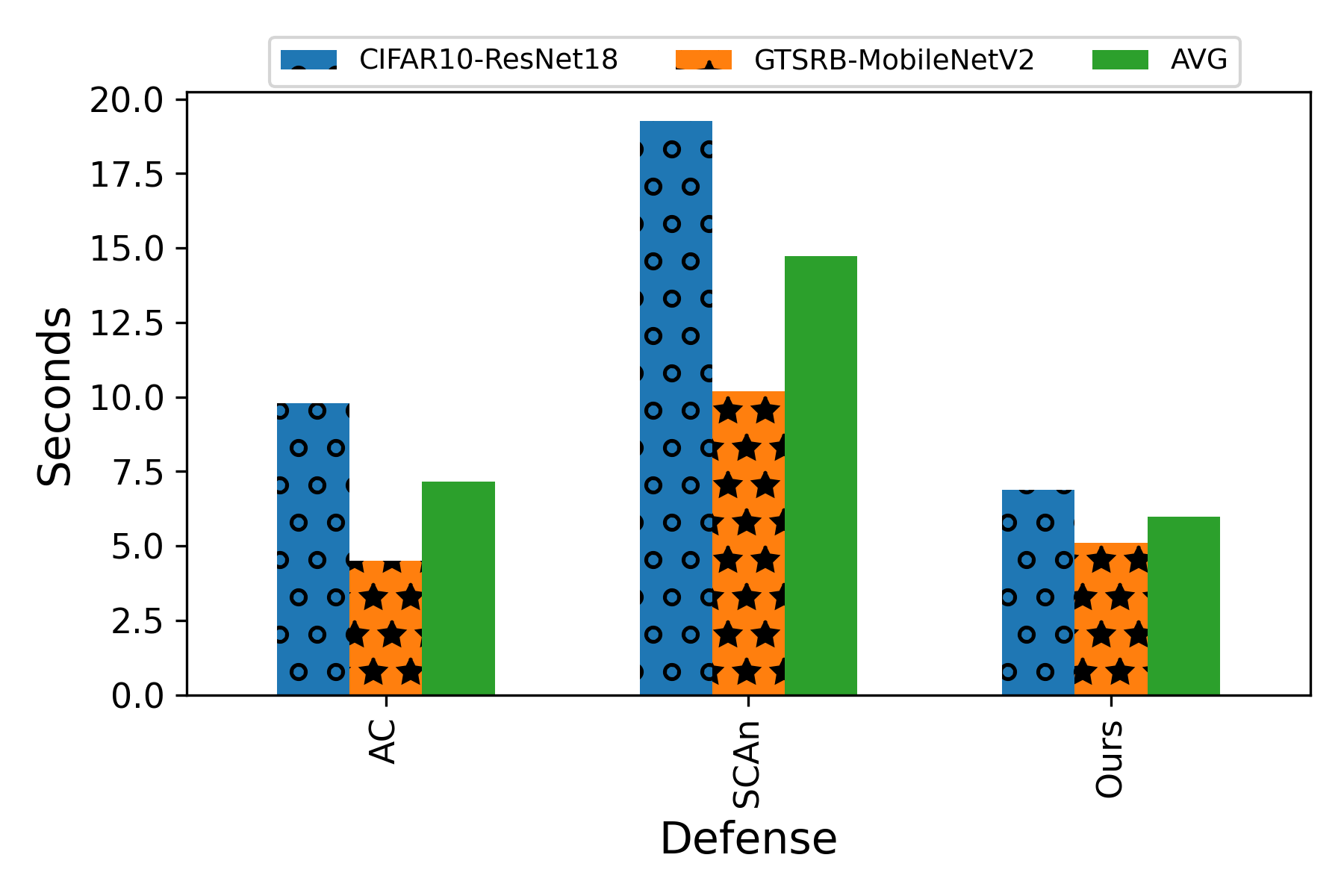}
\caption{Average CPU running time in seconds.}
\label{fig:runtime}
\end{figure*}

We compared the average CPU runtime (in seconds) of our defense with that of AC and SCAn on the whole benign and poisoned test sets.
Figure~\ref{fig:runtime} shows that our defense had the shortest runtime on CIFAR10-ResNet18 and the second shortest on GTSRB-MobileNetV2.
It had a runtime slightly longer than that of AC on GTSRB-MobileNetV2 because MobileNetV2 contains a larger number of intermediate layers, which increases the time required to analyze them.

\section{Detailed Settings for Experiments}
\label{sec_detailed_setting}
We used the PyTorch framework to implement the experiments on an AMD Ryzen 5 3600 6-core CPU with 32 GB RAM, an NVIDIA GTX 1660 GPU, and Windows 10 OS. In addition, we used the BackdoorBox~\cite{li2022backdoorbox} open toolbox for conducting all attacks and re-implemented the other defenses used in our work.

\subsection{Datasets and DNN Architectures}
\label{sec_datasets_dnns}

Table~\ref{tab:data_models} summarizes the statistics of the used datasets and DNNs and the number of benign samples available to the defender.
Note that, for ease of computation, we consider as a layer each convolutional block other than the first convolutional layer and the last fully connected layer.

\begin{table}[t!]
\centering
\caption{Statistics of the used datasets and DNNs}
\label{tab:data_models}
\resizebox{\textwidth}{!}{%
\begin{tabular}{cccccccc}
\toprule
Dataset & Input size & \# Classes & \# Training samples & \#Test samples & \# Available samples & DNN model   & \# Layers \\ \hline
CIFAR-10 & 3x32x32    & 10         & 50,000              & 10,000         & 1,000                & ResNet18    & 10        \\ \hline
GTSRB   & 3x32x32    & 43         & 39,209              & 12,630         & 1,263                & MobileNetV2 & 19        \\ \bottomrule
\end{tabular}%
}
\end{table}

\subsection{Training Setting}
\label{sec_training_setting}
We used the cross-entropy loss and the SGD optimizer with a momentum 0.9 and weight decay  $5\times10^{-4}$ on all benchmarks. 
We used initial learning rates $0.1$ for ResNet18 and $0.01$ for MobileNetV2, and trained models for $200$ epochs. 
The learning rates were decreased by a factor of $10$ at epochs $100$ and $150$, respectively. 
We set the batch size to $128$ and trained all models until they converged.

\subsection{Attack Setting}
\label{sec_attack_setting}
The target class on all datasets was $1$ for BadNets~\cite{gu2019badnets}, the backdoor attack with blended strategy~\cite{chen2017targeted} (Blended), the invisible sample-specific attack~\cite{li2021invisible} (ISSBA), and the input-aware dynamic attack ~\cite{nguyen2020input} (IAD).
The target classes for the label-consistent attack~\cite{turner2019label} and WaNet~\cite{nguyen2020wanet} were $2$ and $0$, respectively, on all datasets.
The trigger patterns of attacks were the same as those presented in the main
paper.
In particular, we set the blended ratio to $\lambda = 0.1$ for the blended attack on all datasets. We used the label-consistent backdoor attack with maximum perturbation size $16$. For WaNet, we set the noise rate to $\rho_n = 0.2$, the control grid size to $k = 4$, and the warping strength to $s = 0.5$ on all datasets, as suggested in the WaNet paper~\cite{nguyen2020wanet}.
For IAD~\cite{nguyen2020input}, we trained the classifier and the trigger generator concurrently. We attached the dynamic trigger to the samples from other classes and relabeled them as the target label.

\subsection{Defense Setting}
\label{sec_defense_setting}
For RS, we generated $100$ neighbors of each input with a mean = $0$ and a standard deviation = $0.1$, as suggested in~\cite{cohen2019certified}.
We set the shrinking rate to $10\%$ for ShPd and padded shrinked images with 0-pixels to expand them to their original size, as suggested in~\cite{li2021backdoor}.
For FP, we pruned $95\%$ of the dormant neurons in the last convolution layer and fine-tuned the pruned model using $5\%$ of the training set.
We adjusted RS, ShPd, and FP to be used as detectors for poisoned samples by comparing the change in prediction before and after applying them to an incoming input.
For AC, STRIP, SCAn, and our defense, we randomly selected $10\%$ from each benign test set as the available benign samples.
Then, for AC, we used the available benign samples, from each class, for normalizing benign and poisoned test samples and identifying potential poisoned clusters.
For STRIP, we blended each input with $100$ random inputs from the available benign samples using a blending value $\alpha = 0.5$, as suggested in~\cite{gao2022design}. 
Then, we identified inputs with entropy below the $10$-th percentile of the entropies of benign samples as potentially poisoned samples.
For SCAn, we identified classes with scores larger than $e$ as potential target classes, as suggested in~\cite{tang2021demon}, and identified the cluster that did not contain the available benign samples as a poisoned cluster.
For our defense, we used a threshold $\tau = 2.5$, which gave us a reasonable trade-off between $TPR$ and $FPR$ on both benchmarks.

\end{document}